\documentclass[aip,apl,reprint,superscriptaddress,10pt]{revtex4-1}

\usepackage{graphicx}

\usepackage{amsmath}

\usepackage{color}

\usepackage{epstopdf}

\usepackage{multirow,bigstrut}

\usepackage{time,mathptmx,helvet}
\usepackage{lineno} 

\usepackage[
pdfstartview=XYZ,
bookmarks=false,
colorlinks=true,
linkcolor=blue,
urlcolor=blue,
citecolor=blue,
pdftex
]{hyperref}
\usepackage{xcolor}
\usepackage{soul}
\makeatletter

\newcommand{\Rmnum}[1]{\expandafter\@slowromancap\romannumeral #1@}
\makeatother

\begin{document}
\title{\large\textcolor{blue}{Record Magnetic Field Generation by Short-Pulse Laser-Driven Capacitor-Coil Targets}\vspace{3pt}}

\author{Lan~Gao}
\altaffiliation{Author to whom correspondence should be addressed: \href{mailto:lgao@pppl.gov}{lgao@pppl.gov}}
\affiliation{Princeton Plasma Physics Laboratory, Princeton University, Princeton, NJ, 08543, USA} 

\author{Yang~Zhang}

\affiliation{Department of Astrophysical Sciences, Princeton University, Princeton, NJ, 08544, USA} 
\affiliation{University Corporation for Atmospheric Research, Boulder, CO, 80301, USA }

\author{Hantao~Ji}
\affiliation{Princeton Plasma Physics Laboratory, Princeton University, Princeton, NJ, 08543, USA} 
\affiliation{Department of Astrophysical Sciences, Princeton University, Princeton, NJ, 08544, USA} 

\author{Brandon~K.~Russell}
\affiliation{Department of Astrophysical Sciences, Princeton University, Princeton, NJ, 08544, USA} 

\author{Geoffrey~Pomraning}
\affiliation{Department of Astrophysical Sciences, Princeton University, Princeton, NJ, 08544, USA} 

\author{Jesse~Griff-McMahon}
\affiliation{Department of Astrophysical Sciences, Princeton University, Princeton, NJ, 08544, USA}

\author{Sallee~Klein}
\affiliation{University of Michigan, Ann Arbor, Michigan, 48109, USA}
 
\author{Carolyn~Kuranz}
\affiliation{University of Michigan, Ann Arbor, Michigan, 48109, USA}

\author{Mingsheng~Wei}
\affiliation{Laboratory for Laser Energetics, University of Rochester, Rochester, New York, 14623, USA}



\begin{abstract}

\noindent Magnetic fields generated by capacitor-coil targets driven by intense short-pulse lasers have been characterized using ultrafast proton radiography. A 1-kJ, 15-ps laser at a center wavelength of 1053 nm irradiated the back plate of the capacitor with an intensity of $\sim$8.3 $\times$ 10$^{18}$ W$/$cm$^{2}$, creating ultra large currents in the connecting coils. High-quality proton data obtained in the axial probing geometry show definitive signatures of magnetic field generation allowing precision measurement of the field distribution and strength. 
The data show a coil current of 120 $\pm$ 10 kA producing 200 $\pm$ 20 Tesla magnetic fields at the coil center at 1.127 ns afer the laser drive. This sets a record for magnetic field generation by the short-pulse-powered capacitor-coil targets.

\end{abstract}




\maketitle

Creation of strong magnetic fields using laser-driven capacitor-coil targets \cite{Daido1986,Courtois2005,fujioka2013kilotesla,Santos_NJP_2015, Gao2016, Law2016, Goyon_PRE_2017, Peebles_PoP_2020, Chien21,Vlachos_PoP_2024,morita2023generation} continues to attract extensive research efforts worldwide due to its potential applications in both basic and applied high-energy-density (HED) science. For a typical design, the target is comprised of two parallel metallic foils connected with a conducting wire. By irradiating one foil with a high-intensity laser, the other becomes negatively charged \cite{PearlmanAPL} collecting superthermal hot electrons generated during the intense laser-solid interaction. \cite{Forslund} This builds up a strong electrical potential between the two foils, \cite{Fiksel_APL_2016} resulting in a large current flowing through the connecting wire and therefore strong magnetic field generation.  

The capacitor-coil target is usually laser cut and bent into the required coil shape, generating a range of magnetic field configurations. The target has an open geometry providing easy access for diagnostic views and magnetizing secondary samples. With a quasi-static magnetic field persisting several ns in $\sim$mm$^{3}$ volumes, lasers-driven capacitor-coil targets have been successfully used in collimating relativistic electrons,\cite{Bailly_NC_2018,Sakata_NC_2018} affecting hydrodynamic instability growth rate,\cite{Matsuo_PRL_2021}, modifying collisionless shock formation \cite{Woolsey_PoP_2001} and hot electron generation in laser-produced plasmas,\cite{Pisarczyk_PPCF_2022} and unraveling physics in magnetic reconnection.\cite{chien2019,chien23,zhang23,ji_PoP_2024} Applications in magnetizing fusion capsule for yield enhancement have also been sought.\cite{Perkins_PoP_2013} 

To push the frontiers of laser-driven capacitor-coil targets for broader applications, stronger magnetic field generation with accurate and reliable magnetic field measurement is essential. The majority of existing work have used ns-scale lasers with intensities in the range of 10$^{14}$ - 10$^{17}$ W$/$cm$^{2}$ to drive the targets reporting magnetic fields from tens to hundreds of Telsa.\cite{Daido1986,Courtois2005,fujioka2013kilotesla,Santos_NJP_2015, Gao2016, Law2016, Goyon_PRE_2017, Peebles_PoP_2020, Chien21,Vlachos_PoP_2024,morita2023generation} In contrast, short-pulse lasers--with pulse duration in the fs to ps range--enabled by recent advancements in laser technology, offer a promising platform for generating even stronger magnetic fields at relativistic intensities exceeding 10$^{18}$ W$/$cm$^{2}$.  At such high intensities, multi-MeV electrons have been measured,\cite{Beg_PoP_1997,YTLi_PRE_2004} potentially generating a much larger electrical potential between the foils and, consequently, producing significantly higher coil currents and stronger magnetic fields when applied to capacitor-coil targets.\cite{Wang_PoP_2023} 

Magnetic pickup probes and optical polarimetry have been used as the primary diagnostics in early experiments where field measurement could only be made at $\sim$mm distances from the coil. This distance was necessary to prevent electromagnetic pulses and fast particles from sabotaging the pickup coil signal or limited by the density region associated with the optical probe beam wavelength. \cite{Santos_NJP_2015} The field strength closer to the coil was theoretically inferred based on a model that describes the magnetic field profile along the probe path, which could lead to overestimation of the total magnetic energy stored in the coil \cite{fujioka2013kilotesla}. Direct measurements of the magnetic fields have been achieved using ultrafast laser-driven proton radiography.\cite{Gao2016,Law2016,Santos_NJP_2015,Goyon_PRE_2017, Peebles_PoP_2020,Chien21,Vlachos_PoP_2024} This diagnostic utilizes high-energy protons to probe through the target,\cite{Gao2012,Gao2013,Gao2015} mapping the spatial distribution and time evolution of the magnetic fields around the coil via proton deflections.\cite{Gao2016} As the protons are deflected by both electric and magnetic fields in the plasmas, the challenge is unambiguous differentiation of magnetic fields from the electric fields in forming the experimentally measured proton radiographs.\cite{Peebles_PoP_2020}


\begin{figure}[t]
\includegraphics{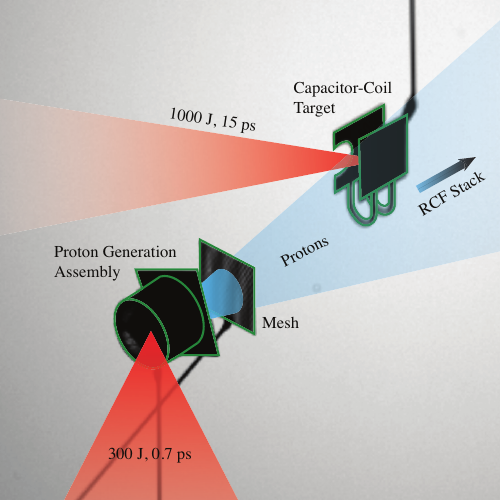}
\caption{\label{fig1} (color online). Experimental setup illustrated with photographic images of the aligned targets (outlined in green), captured by the OMEGA EP target viewing system prior to the actual experiment. The capacitor-coil target consists of two parallel Cu foils connected by two parallel U-shaped Cu coils. The proton generation assembly is composed of a Cu foil mounted inside a plastic tube and a Ta foil attached to the tube end. An Au mesh is positioned 2.5 mm from the Cu foil and 4.5 mm from the center of the U-shaped coils. The Cu foil, Ta foil, and Au mesh are aligned parallel to each other.}
\end{figure}

In this Letter, record magnetic fields generated by capacitor-coil targets driven by intense short-pulse lasers at relativistic laser intensities are reported using axial laser-driven proton radiography. The capacitor-coil target was irradiated by an infrared laser on the OMEGA EP laser system at the University of Rochester's Laboratory for Laser Energetics,\cite{Waxer2005} with $\sim$1-kJ laser energy, 15-ps pulse duration, 16-$\mu$m focal spot radius, and a center wavelength of 1053 nm. This corresponded to a focused laser intensity of $\sim$8.3 $\times$ 10$^{18}$ W$/$cm$^{2}$, which is 2 - 3 orders of magnitude higher than typically used in previous work with ns-long UV drives. \cite{fujioka2013kilotesla,Santos_NJP_2015, Gao2016, Law2016, Goyon_PRE_2017, Peebles_PoP_2020, Chien21,Vlachos_PoP_2024} A multi-MeV proton beam, in combination with a mesh grid, probed through the coil region in the axial geometry (the proton propagation direction is parallel to the coil axis, see Fig.~\ref{fig1}.). High-quality proton data obtained in this setup show definitive signatures that attribute to magnetic field only, allowing precision measurement of the field distribution and strength. The data show a coil current of 120 $\pm$ 10 kA producing 200 $\pm$ 20 Tesla magnetic fields at the coil center at 1.127 ns afer the laser drive. This sets a record for magnetic field generation by the short-pulse-powered capacitor-coil targets and opens up research opportunities in magnetic field generation and application. 

Figure~\ref{fig1} shows a diagram of the experimental setup based on the photographic image of the aligned targets captured by the OMEGA EP target viewing system prior to firing of the lasers. The capacitor-coil target is comprised of two parallel Cu foils (50-$\mu$m thick, 1.5 mm $\times$ 1.5 mm in size, and 600 $\mu$m foil-to-foil separation), connected by two parallel U-shaped Cu coils with a wire cross section 50 $\mu$m $\times$ 50 $\mu$m. Each U-shaped coil has two 500-$\mu$m-long straight wires joined by a half-circular wire with curvature radius of 300 $\mu$m and the inter-coil distance is 600 $\mu$m. An OMEGA EP short-pulse infrared laser beam propagated through a 400-$\mu$m-radius laser entrance hole on the front Cu foil and was focused to the back foil at 45$^{\circ}$ angle of incidence. 

The main interaction was probed with an ultrafast proton beam in a side-on geometry, the so-called axial proton radiography. The proton generation assembly includes a 20-$\mu$m-thick Cu foil and a 5-$\mu$m thick Ta foil that are 1-mm apart. A plastic tube was used to hold both foils, with the Cu foil mounted inside the tube and the Ta foil attached to the side facing the main interaction. Another EP infrared laser (0.3-kJ, 1-ps, and 1053-nm center wavelength) irradiated the Cu foil at 45$^{\circ}$ incidence angle and an intensity of $\sim$1.5 $\times$ 10$^{19}$ W$/$cm$^{2}$, generating tens of MeV protons via the target normal sheath acceleration mechanism (TNSA). \cite{wilks2001} The Ta foil protected the Cu foil from coronal plasma and x-ray photons \cite{zylstra:013511} ensuring a high quality proton beam that provides a spatial resolution of 5 to 10 $\mu$m and a temporal resolution of a few ps.\cite{gao14thesis}

An Au mesh was inserted between the proton assembly and the capacitor-coil target, breaking the proton beam into beamlets. By tracking the deflections of each beamlet as protons pass through the coil region, direct measurements of the fields can be achieved. The Au mesh was a 38-$\mu$m pitch and 26-$\mu$m bar width Au grid placed 2.5 mm away from the Cu foil inside the plastic tube. The protons were eventually detected with a filtered stack of radiochromic film, providing two-dimensional images of the interactions. Each film is related to a proton energy by calculating proton energy deposition inside the film stack at which the Bragg peak occurs, thereby diagnosing the main interaction at times based on the proton time-of-flight to the main target and the timing difference between the drive and proton generation beams. For the experiments reported here, the relative timing between the two short-pulse beams was varied and measured with an x-ray streak camera.\cite{Borghesi_ppcf_ProtonImaging} 

The experimentally measured proton radiographs are magnified proton images with magnification $M$, determined as $M=(D + d)/d$. $d$ is the distance from the proton source foil to the probed region and $D$ is the distance from the probed region to the corresponding radiochromic film with a characteristic proton energy. In these experiments, photographic images of the aligned targets captured by the EP target viewing system were compared with the original design, confirming that the distance from the Cu foil for proton generation was 2.5 mm away from the Au mesh and 7 mm away from the center of coils. The alignment error for the film stack was $\sim$50 $\mu$m. This careful cross-check provided confidence in determining $M$, which was $\sim$ 12.5 to 15 for the coil target and $\sim$ 35 to 42 for the mesh.

\begin{figure}[t]
\includegraphics{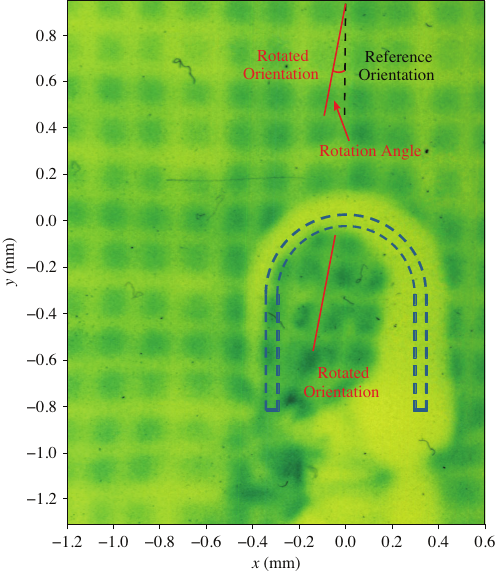}
\caption{\label{fig2} (color online). Proton radiograph obtained with 33 $\pm$ 0.5 MeV protons at $t=t_{0}+1.127$ ns, calibrated to the target plane by accounting for the system magnification. Darker areas correspond to higher detected proton fluxes. The blue dashed line outlines the contour of the coils showing a spatial scale consistent with the experimental setup, further confirming the image calibration. The black dashed line represents the reference mesh grid orientation without any rotation. The red solid line represents the rotated mesh grid orientation in strong magnetic fields. The rotation angle is determined by the tilt of the rotated red solid line relative to the reference black dashed line. Following the coil current from the right leg to the left leg, a distinct width variation is observed.
 }
\end{figure}

Figure~\ref{fig2} shows a typical proton radiograph of the capacitor-coil target after laser irradiation, obtained at $t=t_{0}+1.127$ ns, where $t_{0}$ denotes the arrival time of the short-pulse beam onto the rear Cu plate. The proton energy was 33 $\pm$ 0.5 MeV. The radiograph was calibrated to the target plane by accounting for the system magnification, resulting in a field of view of 1.8 mm $\times$ 2.3 mm, encompassing the entire coil region and a small portion of the foils. The mesh grids outside of the coil region without any distortions were used to cross-check the image calibration, providing an accurate reference for quantitative analyses of distorted region for field inference. Overlaid on top by the blue dashed line is the contour of the coils in the side-on view, showing a spatial scale consistent with the experimental setup. Strong proton deflection is seen below the coil region, likely dominated by the electrical fields in the short-pulse driven plasma plume.  

\begin{figure}[t]
\includegraphics{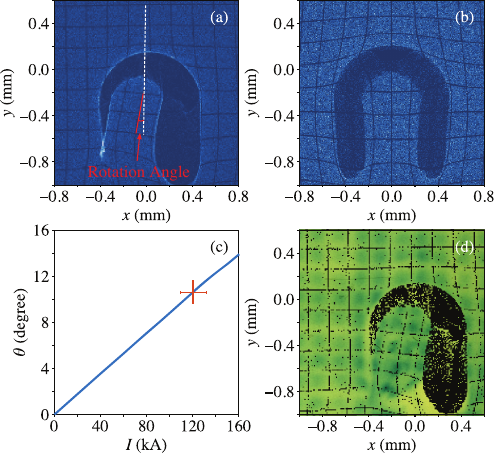}
\caption{\label{fig3} (color online). (a) Synthetic proton radiograph on the target plane calculated with 120 kA current in each coil and no electric fields. Both mesh grid rotation inside the coils and the width variation from the right leg to the left are clearly seen. The white dashed line represents the reference mesh grid orientation without any rotation. The red solid line represents the rotated mesh grid orientation in strong magnetic fields. The rotation angle is determined by the tilt of the rotated red solid line relative to the reference white dashed line. (b) Synthetic proton radiograph calculated from the electric fields produced by a positive line charge within each coil, with a charge density of $+10\;\rm{nC}/\rm{mm}$. (c) The blue solid line represents the simulated rotation angle as a function of coil currents. The red dot indicates the experimentally measured rotation angle of 10.6$^{\circ}$ $\pm$ 1$^{\circ}$ corresponding to a coil current of $120\pm 10$ kA. (d) The trace of the mesh grids and contour of the proton deflection around the coils in (a) is overlaid on top of the measured data as shown in Fig.~\ref{fig2}.}
\end{figure}

Two striking features are observed in the coil region, both of which are caused by magnetic fields generated from the currents flowing in the coils. First, inside the coils, a clear mesh grid rotation in the clockwise direction is observed (see the titled red solid line with respect to the black dashed line in Fig.~\ref{fig2}). As protons propagate through the coils along their axis (into the page in Fig.~\ref{fig2}), they experience a net rotation effect due to non-collinear axial magnetic fields and radial velocities. This feature has been demonstrated to be insensitive to electric fields, \cite{Peebles_PoP_2022} providing robust and unambiguous confirmation of magnetic field generation. Second, a distinct width variation is observed, following the coil current from the right leg to the left leg. This is consistent with the fact that protons are deflected by the azimuthal magnetic fields around the upward-flowing current in the right leg and focused by the downward-flowing current in the left leg. 

The experimental geometry was modeled using the charged particle radiography module of PlasmaPy\cite{plasmapy_community_2024_12788848} to generate synthetic proton radiographs for direct comparison with the measured features. The magnetic field distributions are calculated using the Biot-Savart law for two U-shaped coil currents with coil dimensions taken from the actual target. The electric fields are calculated using uniform charge distributions along each coil mimicking the charge-up effects built in the coils during laser-target interaction. The initial proton-source parameters and radiography geometry were the same as in the experiments. As energetic protons pass through the field region, the proton beam spatial profile undergoes variations due to deflections from the Lorentz force. A synthetic proton image is constructed by tracing each proton trajectory and accumulating proton numbers at the detector plane.

Figure~\ref{fig3}(a) shows a synthetic proton radiograph calculated with 120 kA current in each coil and no electric fields. Within the coils, a clockwise mesh rotation is clearly seen. Along the coil current flowing from the right leg to the left, the transition from strong defocusing to focusing is exhibited. The mesh grids away from the coils are not affected, providing an accurate reference for inferring the rotation angle. The rotation angle is calculated as the tilt of rotated mesh grids with respect to the unrotated mesh grids. The rotation angles inferred from synthetic proton radiographs, similar to the example shown in Fig.~\ref{fig3} (a), are plotted in Fig.~\ref{fig3}(c) for different coil currents, where the rotation angle is found to scale linearly with the coil currents. Such linear relationship is analytically derived in a parallel publication,\cite{zhang2025} providing a theoretical basis for inferring coil currents via the mesh grid rotations.

A similar methodology was used to determine the rotation angle in the experiment, where net apparent rotations of the mesh grids in undisturbed regions--caused by mesh imperfections or alignment errors--was taken into account. With a measured rotation angle of 10.6$^{\circ}$ $\pm$ 1$^{\circ}$, a coil current of $120\pm 10$ kA was inferred.  Fig.~\ref{fig3}(d) shows the synthetic proton image in (a) overlaid on top of the experimental data. All key features, including the mesh rotation, width variation, and mesh grid size, are well reproduced, confirming the inferred coil current at this probing time. The corresponding magnetic field, calculated using the Biot-Savart law, is 200 $\pm$ 20 Tesla at the coil center, setting a record for magnetic field generation by the short-pulse-powered capacitor-coil targets.

In stark contrast to the magnetic field effects, the electric fields deflect protons proportional to the charge distributions in the coils causing the pinching effects for the mesh grids around the coils. There is no mesh rotation or gradual width variation as observed in the case with magnetic fields. Fig.~\ref{fig3}(b) presents an example proton image calculated from the electric fields produced by a positive line charge with charge density of $+10\;{{{\rm{nC}}} \mathord{\left/
 {\vphantom {{{\rm{nC}}} {{\rm{mm}}}}} \right.
 \kern-\nulldelimiterspace} {{\rm{mm}}}}$. A detailed analysis of the electric and magnetic field contributions can be found in Ref.~\citenum{zhang2025}. For the work presented here, contribution of electric fields is negligible, with an upper-bound electric field strength $\sim$3 $\times$ $10^{7}$ V$/$m at 0.3 mm away from the coil.

While record magnetic field strengths are reported here, these do not necessarily represent the peak fields that were produced. Probing the same capacitor-coil target at 3.047 ns after the laser drive, a coil current of $80\pm 10$ kA was measured using the same analysis techniques. Previous work have shown that the coil current keeps growing during the laser irradiation, with a rise time comparable to the laser pulse duration. \cite{Daido1986, Santos_NJP_2015}  After the laser pulse terminates, the target behaves like an inductance-resistance electrical circuit and the current exponentially decays.\cite{Gao2016, Fiksel_APL_2016, Chien21} Assuming a current waveform $I(t) = I_0 e^{- (t - t(0))/\tau}$ where $I_0$ is the peak coil current, $t(0)$ is the laser pulse duration, and $\tau$ is the current decay time, a peak coil current of 150 $\pm$ 20 kA with a decay time of 4.74 $\pm$ 1.75 ns is inferred. Future experiments focused on the early-time dynamics of short-pulse-driven coils are needed to accurately measure the peak field strength.

In summary, direct measurements of the magnetic fields generated by capacitor-coil targets driven by intense short-pulse lasers at relativistic laser intensities are reported. Using the axial proton radiography in combination with a mesh grid, definitive signatures due to magnetic fields are observed. Synthetic proton radiographs calculated with particle ray tracing calculations are compared with the experimentally measured high-quality proton radiographs, matching key signatures of the mesh rotation and coil width evolution. This allows precision measurement of the field distribution and strength. The data show $120 \pm 10$ kA of coil current at 1.127 ns, decaying to $80\pm 10$ kA at 3.047 ns. Magnetic field of 200 $\pm$ 20 Tesla at the coil center is generated, setting a record for magnetic field generation by the short-pulse-powered capacitor-coil targets.


This work opens up research opportunities in magnetic field generation and application at laser facilities that operate exclusively with short-pulse lasers. Recent laser development has increasingly focused on achieving ultra-high intensities, and multipetawatt laser systems have reached intensities four orders of magnitude higher than those used here.\cite{Yoon_2021} Furthermore, there is a significant push to generate stronger fields for applications including novel fusion schemes and studies of plasma processes under conditions relevant to extreme plasma astrophysics.\cite{MP3_workshop}  The coil targets characterized in these experiments are already being used to explore relativistic particle acceleration, enabled by the low-density, high-field environments surrounding the coils.\cite{ji_PoP_2024} The results reported here therefore represent not just an important achievement but a promising path towards even stronger fields and studies at the frontier of HED science.



This work was supported by US Department of Energy the High-Energy-Density Laboratory Plasma Science program under Grant No. DE-SC0020103, the NASA Living with a Star Jack Eddy Postdoctoral Fellowship Program administered by UCAR's Cooperative Programs for the Advancement of Earth System Science (CPAESS) under award $\#$80NSSC22M0097, the LaserNetUS initiative at the OMEGA EP Laser System, and the Laboratory Basic Science program at the Laboratory for Laser Energetics. The authors express their gratitude to OMEGA EP crew for experimental and technical support. 

The data that support the findings of this study are available from the corresponding author upon reasonable request.



\begin{thebibliography}{40}%
\makeatletter
\providecommand \@ifxundefined [1]{%
 \@ifx{#1\undefined}
}%
\providecommand \@ifnum [1]{%
 \ifnum #1\expandafter \@firstoftwo
 \else \expandafter \@secondoftwo
 \fi
}%
\providecommand \@ifx [1]{%
 \ifx #1\expandafter \@firstoftwo
 \else \expandafter \@secondoftwo
 \fi
}%
\providecommand \natexlab [1]{#1}%
\providecommand \enquote  [1]{``#1''}%
\providecommand \bibnamefont  [1]{#1}%
\providecommand \bibfnamefont [1]{#1}%
\providecommand \citenamefont [1]{#1}%
\providecommand \href@noop [0]{\@secondoftwo}%
\providecommand \href [0]{\begingroup \@sanitize@url \@href}%
\providecommand \@href[1]{\@@startlink{#1}\@@href}%
\providecommand \@@href[1]{\endgroup#1\@@endlink}%
\providecommand \@sanitize@url [0]{\catcode `\\12\catcode `\$12\catcode `\&12\catcode `\#12\catcode `\^12\catcode `\_12\catcode `\%12\relax}%
\providecommand \@@startlink[1]{}%
\providecommand \@@endlink[0]{}%
\providecommand \url  [0]{\begingroup\@sanitize@url \@url }%
\providecommand \@url [1]{\endgroup\@href {#1}{\urlprefix }}%
\providecommand \urlprefix  [0]{URL }%
\providecommand \Eprint [0]{\href }%
\providecommand \doibase [0]{http://dx.doi.org/}%
\providecommand \selectlanguage [0]{\@gobble}%
\providecommand \bibinfo  [0]{\@secondoftwo}%
\providecommand \bibfield  [0]{\@secondoftwo}%
\providecommand \translation [1]{[#1]}%
\providecommand \BibitemOpen [0]{}%
\providecommand \bibitemStop [0]{}%
\providecommand \bibitemNoStop [0]{.\EOS\space}%
\providecommand \EOS [0]{\spacefactor3000\relax}%
\providecommand \BibitemShut  [1]{\csname bibitem#1\endcsname}%
\let\auto@bib@innerbib\@empty
\bibitem [{\citenamefont {Daido}\ \emph {et~al.}(1986)\citenamefont {Daido}, \citenamefont {Miki}, \citenamefont {Mima}, \citenamefont {Fujita}, \citenamefont {Sawai}, \citenamefont {Fujita}, \citenamefont {Kitagawa}, \citenamefont {Nakai},\ and\ \citenamefont {Yamanaka}}]{Daido1986}%
  \BibitemOpen
  \bibfield  {author} {\bibinfo {author} {\bibfnamefont {H.}~\bibnamefont {Daido}}, \bibinfo {author} {\bibfnamefont {F.}~\bibnamefont {Miki}}, \bibinfo {author} {\bibfnamefont {K.}~\bibnamefont {Mima}}, \bibinfo {author} {\bibfnamefont {M.}~\bibnamefont {Fujita}}, \bibinfo {author} {\bibfnamefont {K.}~\bibnamefont {Sawai}}, \bibinfo {author} {\bibfnamefont {H.}~\bibnamefont {Fujita}}, \bibinfo {author} {\bibfnamefont {Y.}~\bibnamefont {Kitagawa}}, \bibinfo {author} {\bibfnamefont {S.}~\bibnamefont {Nakai}}, \ and\ \bibinfo {author} {\bibfnamefont {C.}~\bibnamefont {Yamanaka}},\ }\href {\doibase 10.1103/PhysRevLett.56.846} {\bibfield  {journal} {\bibinfo  {journal} {Phys. Rev. Lett.}\ }\textbf {\bibinfo {volume} {56}},\ \bibinfo {pages} {846} (\bibinfo {year} {1986})}\BibitemShut {NoStop}%
\bibitem [{\citenamefont {Courtois}\ \emph {et~al.}(2005)\citenamefont {Courtois}, \citenamefont {Ash}, \citenamefont {Chambers}, \citenamefont {Grundy},\ and\ \citenamefont {Woolsey}}]{Courtois2005}%
  \BibitemOpen
  \bibfield  {author} {\bibinfo {author} {\bibfnamefont {C.}~\bibnamefont {Courtois}}, \bibinfo {author} {\bibfnamefont {A.~D.}\ \bibnamefont {Ash}}, \bibinfo {author} {\bibfnamefont {D.~M.}\ \bibnamefont {Chambers}}, \bibinfo {author} {\bibfnamefont {R.~A.~D.}\ \bibnamefont {Grundy}}, \ and\ \bibinfo {author} {\bibfnamefont {N.~C.}\ \bibnamefont {Woolsey}},\ }\href {\doibase 10.1063/1.2035896} {\bibfield  {journal} {\bibinfo  {journal} {J. Appl. Phys.}\ }\textbf {\bibinfo {volume} {98}},\ \bibinfo {eid} {054913} (\bibinfo {year} {2005})}\BibitemShut {NoStop}%
\bibitem [{\citenamefont {Fujioka}\ \emph {et~al.}(2013)\citenamefont {Fujioka}, \citenamefont {Zhang}, \citenamefont {Ishihara}, \citenamefont {Shigemori}, \citenamefont {Hironaka}, \citenamefont {Johzaki}, \citenamefont {Sunahara}, \citenamefont {Yamamoto}, \citenamefont {Nakashima}, \citenamefont {Watanabe}, \citenamefont {Shiraga}, \citenamefont {Nishimura},\ and\ \citenamefont {Azechi}}]{fujioka2013kilotesla}%
  \BibitemOpen
  \bibfield  {author} {\bibinfo {author} {\bibfnamefont {S.}~\bibnamefont {Fujioka}}, \bibinfo {author} {\bibfnamefont {Z.}~\bibnamefont {Zhang}}, \bibinfo {author} {\bibfnamefont {K.}~\bibnamefont {Ishihara}}, \bibinfo {author} {\bibfnamefont {K.}~\bibnamefont {Shigemori}}, \bibinfo {author} {\bibfnamefont {Y.}~\bibnamefont {Hironaka}}, \bibinfo {author} {\bibfnamefont {T.}~\bibnamefont {Johzaki}}, \bibinfo {author} {\bibfnamefont {A.}~\bibnamefont {Sunahara}}, \bibinfo {author} {\bibfnamefont {N.}~\bibnamefont {Yamamoto}}, \bibinfo {author} {\bibfnamefont {H.}~\bibnamefont {Nakashima}}, \bibinfo {author} {\bibfnamefont {T.}~\bibnamefont {Watanabe}}, \bibinfo {author} {\bibfnamefont {H.}~\bibnamefont {Shiraga}}, \bibinfo {author} {\bibfnamefont {H.}~\bibnamefont {Nishimura}}, \ and\ \bibinfo {author} {\bibfnamefont {H.}~\bibnamefont {Azechi}},\ }\href {http://dx.doi.org/10.1038/srep01170} {\bibfield  {journal} {\bibinfo  {journal} {Sci. Rep.}\ }\textbf {\bibinfo {volume} {3}} (\bibinfo {year}
  {2013})}\BibitemShut {NoStop}%
\bibitem [{\citenamefont {Santos}\ \emph {et~al.}(2015)\citenamefont {Santos}, \citenamefont {Bailly-Grandvaux}, \citenamefont {Giuffrida}, \citenamefont {Forestier-Colleoni}, \citenamefont {Fujioka}, \citenamefont {Zhang}, \citenamefont {Korneev}, \citenamefont {Bouillaud}, \citenamefont {Dorard}, \citenamefont {Batani}, \citenamefont {Chevrot}, \citenamefont {Cross}, \citenamefont {Crowston}, \citenamefont {Dubois}, \citenamefont {Gazave}, \citenamefont {Gregori}, \citenamefont {d’Humières}, \citenamefont {Hulin}, \citenamefont {Ishihara}, \citenamefont {Kojima}, \citenamefont {Loyez}, \citenamefont {Marquès}, \citenamefont {Morace}, \citenamefont {Nicolaï}, \citenamefont {Peyrusse}, \citenamefont {Poyé}, \citenamefont {Raffestin}, \citenamefont {Ribolzi}, \citenamefont {Roth}, \citenamefont {Schaumann}, \citenamefont {Serres}, \citenamefont {Tikhonchuk}, \citenamefont {Vacar},\ and\ \citenamefont {Woolsey}}]{Santos_NJP_2015}%
  \BibitemOpen
  \bibfield  {author} {\bibinfo {author} {\bibfnamefont {J.~J.}\ \bibnamefont {Santos}}, \bibinfo {author} {\bibfnamefont {M.}~\bibnamefont {Bailly-Grandvaux}}, \bibinfo {author} {\bibfnamefont {L.}~\bibnamefont {Giuffrida}}, \bibinfo {author} {\bibfnamefont {P.}~\bibnamefont {Forestier-Colleoni}}, \bibinfo {author} {\bibfnamefont {S.}~\bibnamefont {Fujioka}}, \bibinfo {author} {\bibfnamefont {Z.}~\bibnamefont {Zhang}}, \bibinfo {author} {\bibfnamefont {P.}~\bibnamefont {Korneev}}, \bibinfo {author} {\bibfnamefont {R.}~\bibnamefont {Bouillaud}}, \bibinfo {author} {\bibfnamefont {S.}~\bibnamefont {Dorard}}, \bibinfo {author} {\bibfnamefont {D.}~\bibnamefont {Batani}}, \bibinfo {author} {\bibfnamefont {M.}~\bibnamefont {Chevrot}}, \bibinfo {author} {\bibfnamefont {J.~E.}\ \bibnamefont {Cross}}, \bibinfo {author} {\bibfnamefont {R.}~\bibnamefont {Crowston}}, \bibinfo {author} {\bibfnamefont {J.-L.}\ \bibnamefont {Dubois}}, \bibinfo {author} {\bibfnamefont {J.}~\bibnamefont {Gazave}}, \bibinfo {author}
  {\bibfnamefont {G.}~\bibnamefont {Gregori}}, \bibinfo {author} {\bibfnamefont {E.}~\bibnamefont {d’Humières}}, \bibinfo {author} {\bibfnamefont {S.}~\bibnamefont {Hulin}}, \bibinfo {author} {\bibfnamefont {K.}~\bibnamefont {Ishihara}}, \bibinfo {author} {\bibfnamefont {S.}~\bibnamefont {Kojima}}, \bibinfo {author} {\bibfnamefont {E.}~\bibnamefont {Loyez}}, \bibinfo {author} {\bibfnamefont {J.-R.}\ \bibnamefont {Marquès}}, \bibinfo {author} {\bibfnamefont {A.}~\bibnamefont {Morace}}, \bibinfo {author} {\bibfnamefont {P.}~\bibnamefont {Nicolaï}}, \bibinfo {author} {\bibfnamefont {O.}~\bibnamefont {Peyrusse}}, \bibinfo {author} {\bibfnamefont {A.}~\bibnamefont {Poyé}}, \bibinfo {author} {\bibfnamefont {D.}~\bibnamefont {Raffestin}}, \bibinfo {author} {\bibfnamefont {J.}~\bibnamefont {Ribolzi}}, \bibinfo {author} {\bibfnamefont {M.}~\bibnamefont {Roth}}, \bibinfo {author} {\bibfnamefont {G.}~\bibnamefont {Schaumann}}, \bibinfo {author} {\bibfnamefont {F.}~\bibnamefont {Serres}}, \bibinfo {author}
  {\bibfnamefont {V.~T.}\ \bibnamefont {Tikhonchuk}}, \bibinfo {author} {\bibfnamefont {P.}~\bibnamefont {Vacar}}, \ and\ \bibinfo {author} {\bibfnamefont {N.}~\bibnamefont {Woolsey}},\ }\href {\doibase 10.1088/1367-2630/17/8/083051} {\bibfield  {journal} {\bibinfo  {journal} {New Journal of Physics}\ }\textbf {\bibinfo {volume} {17}},\ \bibinfo {pages} {083051} (\bibinfo {year} {2015})}\BibitemShut {NoStop}%
\bibitem [{\citenamefont {Gao}\ \emph {et~al.}(2016)\citenamefont {Gao}, \citenamefont {Ji}, \citenamefont {Fiksel}, \citenamefont {Fox}, \citenamefont {Evans},\ and\ \citenamefont {Alfonso}}]{Gao2016}%
  \BibitemOpen
  \bibfield  {author} {\bibinfo {author} {\bibfnamefont {L.}~\bibnamefont {Gao}}, \bibinfo {author} {\bibfnamefont {H.}~\bibnamefont {Ji}}, \bibinfo {author} {\bibfnamefont {G.}~\bibnamefont {Fiksel}}, \bibinfo {author} {\bibfnamefont {W.}~\bibnamefont {Fox}}, \bibinfo {author} {\bibfnamefont {M.}~\bibnamefont {Evans}}, \ and\ \bibinfo {author} {\bibfnamefont {N.}~\bibnamefont {Alfonso}},\ }\href {\doibase 10.1063/1.4945643} {\bibfield  {journal} {\bibinfo  {journal} {Physics of Plasmas}\ }\textbf {\bibinfo {volume} {23}},\ \bibinfo {pages} {043106} (\bibinfo {year} {2016})}\BibitemShut {NoStop}%
\bibitem [{\citenamefont {Law}\ \emph {et~al.}(2016)\citenamefont {Law}, \citenamefont {Bailly-Grandvaux}, \citenamefont {Morace}, \citenamefont {Sakata}, \citenamefont {Matsuo}, \citenamefont {Kojima}, \citenamefont {Lee}, \citenamefont {Vaisseau}, \citenamefont {Arikawa}, \citenamefont {Yogo}, \citenamefont {Kondo}, \citenamefont {Zhang}, \citenamefont {Bellei}, \citenamefont {Santos}, \citenamefont {Fujioka},\ and\ \citenamefont {Azechi}}]{Law2016}%
  \BibitemOpen
  \bibfield  {author} {\bibinfo {author} {\bibfnamefont {K.~F.~F.}\ \bibnamefont {Law}}, \bibinfo {author} {\bibfnamefont {M.}~\bibnamefont {Bailly-Grandvaux}}, \bibinfo {author} {\bibfnamefont {A.}~\bibnamefont {Morace}}, \bibinfo {author} {\bibfnamefont {S.}~\bibnamefont {Sakata}}, \bibinfo {author} {\bibfnamefont {K.}~\bibnamefont {Matsuo}}, \bibinfo {author} {\bibfnamefont {S.}~\bibnamefont {Kojima}}, \bibinfo {author} {\bibfnamefont {S.}~\bibnamefont {Lee}}, \bibinfo {author} {\bibfnamefont {X.}~\bibnamefont {Vaisseau}}, \bibinfo {author} {\bibfnamefont {Y.}~\bibnamefont {Arikawa}}, \bibinfo {author} {\bibfnamefont {A.}~\bibnamefont {Yogo}}, \bibinfo {author} {\bibfnamefont {K.}~\bibnamefont {Kondo}}, \bibinfo {author} {\bibfnamefont {Z.}~\bibnamefont {Zhang}}, \bibinfo {author} {\bibfnamefont {C.}~\bibnamefont {Bellei}}, \bibinfo {author} {\bibfnamefont {J.~J.}\ \bibnamefont {Santos}}, \bibinfo {author} {\bibfnamefont {S.}~\bibnamefont {Fujioka}}, \ and\ \bibinfo {author} {\bibfnamefont
  {H.}~\bibnamefont {Azechi}},\ }\href {\doibase 10.1063/1.4943078} {\bibfield  {journal} {\bibinfo  {journal} {Applied Physics Letters}\ }\textbf {\bibinfo {volume} {108}},\ \bibinfo {pages} {091104} (\bibinfo {year} {2016})}\BibitemShut {NoStop}%
\bibitem [{\citenamefont {Goyon}\ \emph {et~al.}(2017)\citenamefont {Goyon}, \citenamefont {Pollock}, \citenamefont {Turnbull}, \citenamefont {Hazi}, \citenamefont {Divol}, \citenamefont {Farmer}, \citenamefont {Haberberger}, \citenamefont {Javedani}, \citenamefont {Johnson}, \citenamefont {Kemp}, \citenamefont {Levy}, \citenamefont {Grant~Logan}, \citenamefont {Mariscal}, \citenamefont {Landen}, \citenamefont {Patankar}, \citenamefont {Ross}, \citenamefont {Rubenchik}, \citenamefont {Swadling}, \citenamefont {Williams}, \citenamefont {Fujioka}, \citenamefont {Law},\ and\ \citenamefont {Moody}}]{Goyon_PRE_2017}%
  \BibitemOpen
  \bibfield  {author} {\bibinfo {author} {\bibfnamefont {C.}~\bibnamefont {Goyon}}, \bibinfo {author} {\bibfnamefont {B.~B.}\ \bibnamefont {Pollock}}, \bibinfo {author} {\bibfnamefont {D.~P.}\ \bibnamefont {Turnbull}}, \bibinfo {author} {\bibfnamefont {A.}~\bibnamefont {Hazi}}, \bibinfo {author} {\bibfnamefont {L.}~\bibnamefont {Divol}}, \bibinfo {author} {\bibfnamefont {W.~A.}\ \bibnamefont {Farmer}}, \bibinfo {author} {\bibfnamefont {D.}~\bibnamefont {Haberberger}}, \bibinfo {author} {\bibfnamefont {J.}~\bibnamefont {Javedani}}, \bibinfo {author} {\bibfnamefont {A.~J.}\ \bibnamefont {Johnson}}, \bibinfo {author} {\bibfnamefont {A.}~\bibnamefont {Kemp}}, \bibinfo {author} {\bibfnamefont {M.~C.}\ \bibnamefont {Levy}}, \bibinfo {author} {\bibfnamefont {B.}~\bibnamefont {Grant~Logan}}, \bibinfo {author} {\bibfnamefont {D.~A.}\ \bibnamefont {Mariscal}}, \bibinfo {author} {\bibfnamefont {O.~L.}\ \bibnamefont {Landen}}, \bibinfo {author} {\bibfnamefont {S.}~\bibnamefont {Patankar}}, \bibinfo {author}
  {\bibfnamefont {J.~S.}\ \bibnamefont {Ross}}, \bibinfo {author} {\bibfnamefont {A.~M.}\ \bibnamefont {Rubenchik}}, \bibinfo {author} {\bibfnamefont {G.~F.}\ \bibnamefont {Swadling}}, \bibinfo {author} {\bibfnamefont {G.~J.}\ \bibnamefont {Williams}}, \bibinfo {author} {\bibfnamefont {S.}~\bibnamefont {Fujioka}}, \bibinfo {author} {\bibfnamefont {K.~F.~F.}\ \bibnamefont {Law}}, \ and\ \bibinfo {author} {\bibfnamefont {J.~D.}\ \bibnamefont {Moody}},\ }\href {\doibase 10.1103/PhysRevE.95.033208} {\bibfield  {journal} {\bibinfo  {journal} {Phys. Rev. E}\ }\textbf {\bibinfo {volume} {95}},\ \bibinfo {pages} {033208} (\bibinfo {year} {2017})}\BibitemShut {NoStop}%
\bibitem [{\citenamefont {Peebles}\ \emph {et~al.}(2020)\citenamefont {Peebles}, \citenamefont {Davies}, \citenamefont {Barnak}, \citenamefont {Cracium}, \citenamefont {Bonino},\ and\ \citenamefont {Betti}}]{Peebles_PoP_2020}%
  \BibitemOpen
  \bibfield  {author} {\bibinfo {author} {\bibfnamefont {J.~L.}\ \bibnamefont {Peebles}}, \bibinfo {author} {\bibfnamefont {J.~R.}\ \bibnamefont {Davies}}, \bibinfo {author} {\bibfnamefont {D.~H.}\ \bibnamefont {Barnak}}, \bibinfo {author} {\bibfnamefont {T.}~\bibnamefont {Cracium}}, \bibinfo {author} {\bibfnamefont {M.~J.}\ \bibnamefont {Bonino}}, \ and\ \bibinfo {author} {\bibfnamefont {R.}~\bibnamefont {Betti}},\ }\href {\doibase 10.1063/1.5134786} {\bibfield  {journal} {\bibinfo  {journal} {Physics of Plasmas}\ }\textbf {\bibinfo {volume} {27}},\ \bibinfo {pages} {063109} (\bibinfo {year} {2020})}\BibitemShut {NoStop}%
\bibitem [{\citenamefont {Chien}\ \emph {et~al.}(2021)\citenamefont {Chien}, \citenamefont {Gao}, \citenamefont {Zhang}, \citenamefont {Ji}, \citenamefont {Blackman}, \citenamefont {Chen}, \citenamefont {Fiksel}, \citenamefont {Hill},\ and\ \citenamefont {Nilson}}]{Chien21}%
  \BibitemOpen
  \bibfield  {author} {\bibinfo {author} {\bibfnamefont {A.}~\bibnamefont {Chien}}, \bibinfo {author} {\bibfnamefont {L.}~\bibnamefont {Gao}}, \bibinfo {author} {\bibfnamefont {S.}~\bibnamefont {Zhang}}, \bibinfo {author} {\bibfnamefont {H.}~\bibnamefont {Ji}}, \bibinfo {author} {\bibfnamefont {E.}~\bibnamefont {Blackman}}, \bibinfo {author} {\bibfnamefont {H.}~\bibnamefont {Chen}}, \bibinfo {author} {\bibfnamefont {G.}~\bibnamefont {Fiksel}}, \bibinfo {author} {\bibfnamefont {K.}~\bibnamefont {Hill}}, \ and\ \bibinfo {author} {\bibfnamefont {P.}~\bibnamefont {Nilson}},\ }\href {\doibase 10.1063/5.0044048} {\bibfield  {journal} {\bibinfo  {journal} {Physics of Plasmas}\ }\textbf {\bibinfo {volume} {28}},\ \bibinfo {pages} {052105} (\bibinfo {year} {2021})}\BibitemShut {NoStop}%
\bibitem [{\citenamefont {Vlachos}\ \emph {et~al.}(2024)\citenamefont {Vlachos}, \citenamefont {Ospina-Bohórquez}, \citenamefont {Bradford}, \citenamefont {Pérez-Callejo}, \citenamefont {Ehret}, \citenamefont {Guillon}, \citenamefont {Lendrin}, \citenamefont {Vaisseau}, \citenamefont {Albertazzi}, \citenamefont {Soussan}, \citenamefont {Koenig}, \citenamefont {Malko}, \citenamefont {Kaur}, \citenamefont {Gjevre}, \citenamefont {Fedosejevs}, \citenamefont {Bailly-Grandvaux}, \citenamefont {Walsh}, \citenamefont {Florido}, \citenamefont {Suzuki-Vidal}, \citenamefont {McGuffey}, \citenamefont {Saret}, \citenamefont {Beg}, \citenamefont {Chodukowski}, \citenamefont {Pisarczyk}, \citenamefont {Rusiniak}, \citenamefont {Dostal}, \citenamefont {Dudzak}, \citenamefont {Calisti}, \citenamefont {Ferri}, \citenamefont {Volpe}, \citenamefont {Woolsey}, \citenamefont {Gremillet}, \citenamefont {Tikhonchuk},\ and\ \citenamefont {Santos}}]{Vlachos_PoP_2024}%
  \BibitemOpen
  \bibfield  {author} {\bibinfo {author} {\bibfnamefont {C.}~\bibnamefont {Vlachos}}, \bibinfo {author} {\bibfnamefont {V.}~\bibnamefont {Ospina-Bohórquez}}, \bibinfo {author} {\bibfnamefont {P.~W.}\ \bibnamefont {Bradford}}, \bibinfo {author} {\bibfnamefont {G.}~\bibnamefont {Pérez-Callejo}}, \bibinfo {author} {\bibfnamefont {M.}~\bibnamefont {Ehret}}, \bibinfo {author} {\bibfnamefont {P.}~\bibnamefont {Guillon}}, \bibinfo {author} {\bibfnamefont {M.}~\bibnamefont {Lendrin}}, \bibinfo {author} {\bibfnamefont {X.}~\bibnamefont {Vaisseau}}, \bibinfo {author} {\bibfnamefont {B.}~\bibnamefont {Albertazzi}}, \bibinfo {author} {\bibfnamefont {E.}~\bibnamefont {Soussan}}, \bibinfo {author} {\bibfnamefont {M.}~\bibnamefont {Koenig}}, \bibinfo {author} {\bibfnamefont {S.}~\bibnamefont {Malko}}, \bibinfo {author} {\bibfnamefont {C.}~\bibnamefont {Kaur}}, \bibinfo {author} {\bibfnamefont {M.}~\bibnamefont {Gjevre}}, \bibinfo {author} {\bibfnamefont {R.}~\bibnamefont {Fedosejevs}}, \bibinfo {author} {\bibfnamefont
  {M.}~\bibnamefont {Bailly-Grandvaux}}, \bibinfo {author} {\bibfnamefont {C.~A.}\ \bibnamefont {Walsh}}, \bibinfo {author} {\bibfnamefont {R.}~\bibnamefont {Florido}}, \bibinfo {author} {\bibfnamefont {F.}~\bibnamefont {Suzuki-Vidal}}, \bibinfo {author} {\bibfnamefont {C.}~\bibnamefont {McGuffey}}, \bibinfo {author} {\bibfnamefont {J.}~\bibnamefont {Saret}}, \bibinfo {author} {\bibfnamefont {F.~N.}\ \bibnamefont {Beg}}, \bibinfo {author} {\bibfnamefont {T.}~\bibnamefont {Chodukowski}}, \bibinfo {author} {\bibfnamefont {T.}~\bibnamefont {Pisarczyk}}, \bibinfo {author} {\bibfnamefont {Z.}~\bibnamefont {Rusiniak}}, \bibinfo {author} {\bibfnamefont {J.}~\bibnamefont {Dostal}}, \bibinfo {author} {\bibfnamefont {R.}~\bibnamefont {Dudzak}}, \bibinfo {author} {\bibfnamefont {A.}~\bibnamefont {Calisti}}, \bibinfo {author} {\bibfnamefont {S.}~\bibnamefont {Ferri}}, \bibinfo {author} {\bibfnamefont {L.}~\bibnamefont {Volpe}}, \bibinfo {author} {\bibfnamefont {N.~C.}\ \bibnamefont {Woolsey}}, \bibinfo {author}
  {\bibfnamefont {L.}~\bibnamefont {Gremillet}}, \bibinfo {author} {\bibfnamefont {V.}~\bibnamefont {Tikhonchuk}}, \ and\ \bibinfo {author} {\bibfnamefont {J.~J.}\ \bibnamefont {Santos}},\ }\href {\doibase 10.1063/5.0190305} {\bibfield  {journal} {\bibinfo  {journal} {Physics of Plasmas}\ }\textbf {\bibinfo {volume} {31}},\ \bibinfo {pages} {032702} (\bibinfo {year} {2024})}\BibitemShut {NoStop}%
\bibitem [{\citenamefont {Morita}\ and\ \citenamefont {Fujioka}(2023)}]{morita2023generation}%
  \BibitemOpen
  \bibfield  {author} {\bibinfo {author} {\bibfnamefont {H.}~\bibnamefont {Morita}}\ and\ \bibinfo {author} {\bibfnamefont {S.}~\bibnamefont {Fujioka}},\ }\href@noop {} {\bibfield  {journal} {\bibinfo  {journal} {Reviews of Modern Plasma Physics}\ }\textbf {\bibinfo {volume} {7}},\ \bibinfo {pages} {13} (\bibinfo {year} {2023})}\BibitemShut {NoStop}%
\bibitem [{\citenamefont {Pearlman}\ and\ \citenamefont {Dahlbacka}(1977)}]{PearlmanAPL}%
  \BibitemOpen
  \bibfield  {author} {\bibinfo {author} {\bibfnamefont {J.~S.}\ \bibnamefont {Pearlman}}\ and\ \bibinfo {author} {\bibfnamefont {G.~H.}\ \bibnamefont {Dahlbacka}},\ }\href {\doibase 10.1063/1.89729} {\bibfield  {journal} {\bibinfo  {journal} {Appl. Phys. Lett.}\ }\textbf {\bibinfo {volume} {31}},\ \bibinfo {pages} {414} (\bibinfo {year} {1977})}\BibitemShut {NoStop}%
\bibitem [{\citenamefont {Forslund}, \citenamefont {Kindel},\ and\ \citenamefont {Lee}(1977)}]{Forslund}%
  \BibitemOpen
  \bibfield  {author} {\bibinfo {author} {\bibfnamefont {D.~W.}\ \bibnamefont {Forslund}}, \bibinfo {author} {\bibfnamefont {J.~M.}\ \bibnamefont {Kindel}}, \ and\ \bibinfo {author} {\bibfnamefont {K.}~\bibnamefont {Lee}},\ }\href {\doibase 10.1103/PhysRevLett.39.284} {\bibfield  {journal} {\bibinfo  {journal} {Phys. Rev. Lett.}\ }\textbf {\bibinfo {volume} {39}},\ \bibinfo {pages} {284} (\bibinfo {year} {1977})}\BibitemShut {NoStop}%
\bibitem [{\citenamefont {Fiksel}\ \emph {et~al.}(2016)\citenamefont {Fiksel}, \citenamefont {Fox}, \citenamefont {Gao},\ and\ \citenamefont {Ji}}]{Fiksel_APL_2016}%
  \BibitemOpen
  \bibfield  {author} {\bibinfo {author} {\bibfnamefont {G.}~\bibnamefont {Fiksel}}, \bibinfo {author} {\bibfnamefont {W.}~\bibnamefont {Fox}}, \bibinfo {author} {\bibfnamefont {L.}~\bibnamefont {Gao}}, \ and\ \bibinfo {author} {\bibfnamefont {H.}~\bibnamefont {Ji}},\ }\href {\doibase 10.1063/1.4963763} {\bibfield  {journal} {\bibinfo  {journal} {Applied Physics Letters}\ }\textbf {\bibinfo {volume} {109}},\ \bibinfo {pages} {134103} (\bibinfo {year} {2016})}\BibitemShut {NoStop}%
\bibitem [{\citenamefont {Bailly-Grandvaux}\ \emph {et~al.}(2018)\citenamefont {Bailly-Grandvaux}, \citenamefont {Santos}, \citenamefont {Bellei}, \citenamefont {Forestier-Colleoni}, \citenamefont {Fujioka}, \citenamefont {Giuffrida}, \citenamefont {Honrubia}, \citenamefont {Batani}, \citenamefont {Bouillaud}, \citenamefont {Chevrot}, \citenamefont {Cross}, \citenamefont {Crowston}, \citenamefont {Dorard}, \citenamefont {Dubois}, \citenamefont {Ehret}, \citenamefont {Gregori}, \citenamefont {Hulin}, \citenamefont {Kojima}, \citenamefont {Loyez}, \citenamefont {Marquès}, \citenamefont {Morace}, \citenamefont {Nicolaï}, \citenamefont {Roth}, \citenamefont {Sakata}, \citenamefont {Schaumann}, \citenamefont {Serres}, \citenamefont {Servel}, \citenamefont {Tikhonchuk}, \citenamefont {Woolsey},\ and\ \citenamefont {Zhang}}]{Bailly_NC_2018}%
  \BibitemOpen
  \bibfield  {author} {\bibinfo {author} {\bibfnamefont {M.}~\bibnamefont {Bailly-Grandvaux}}, \bibinfo {author} {\bibfnamefont {J.~J.}\ \bibnamefont {Santos}}, \bibinfo {author} {\bibfnamefont {C.}~\bibnamefont {Bellei}}, \bibinfo {author} {\bibfnamefont {P.}~\bibnamefont {Forestier-Colleoni}}, \bibinfo {author} {\bibfnamefont {S.}~\bibnamefont {Fujioka}}, \bibinfo {author} {\bibfnamefont {L.}~\bibnamefont {Giuffrida}}, \bibinfo {author} {\bibfnamefont {J.~J.}\ \bibnamefont {Honrubia}}, \bibinfo {author} {\bibfnamefont {D.}~\bibnamefont {Batani}}, \bibinfo {author} {\bibfnamefont {R.}~\bibnamefont {Bouillaud}}, \bibinfo {author} {\bibfnamefont {M.}~\bibnamefont {Chevrot}}, \bibinfo {author} {\bibfnamefont {J.~E.}\ \bibnamefont {Cross}}, \bibinfo {author} {\bibfnamefont {R.}~\bibnamefont {Crowston}}, \bibinfo {author} {\bibfnamefont {S.}~\bibnamefont {Dorard}}, \bibinfo {author} {\bibfnamefont {J.~L.}\ \bibnamefont {Dubois}}, \bibinfo {author} {\bibfnamefont {M.}~\bibnamefont {Ehret}}, \bibinfo {author}
  {\bibfnamefont {G.}~\bibnamefont {Gregori}}, \bibinfo {author} {\bibfnamefont {S.}~\bibnamefont {Hulin}}, \bibinfo {author} {\bibfnamefont {S.}~\bibnamefont {Kojima}}, \bibinfo {author} {\bibfnamefont {E.}~\bibnamefont {Loyez}}, \bibinfo {author} {\bibfnamefont {J.~R.}\ \bibnamefont {Marquès}}, \bibinfo {author} {\bibfnamefont {A.}~\bibnamefont {Morace}}, \bibinfo {author} {\bibfnamefont {P.}~\bibnamefont {Nicolaï}}, \bibinfo {author} {\bibfnamefont {M.}~\bibnamefont {Roth}}, \bibinfo {author} {\bibfnamefont {S.}~\bibnamefont {Sakata}}, \bibinfo {author} {\bibfnamefont {G.}~\bibnamefont {Schaumann}}, \bibinfo {author} {\bibfnamefont {F.}~\bibnamefont {Serres}}, \bibinfo {author} {\bibfnamefont {J.}~\bibnamefont {Servel}}, \bibinfo {author} {\bibfnamefont {V.~T.}\ \bibnamefont {Tikhonchuk}}, \bibinfo {author} {\bibfnamefont {N.}~\bibnamefont {Woolsey}}, \ and\ \bibinfo {author} {\bibfnamefont {Z.}~\bibnamefont {Zhang}},\ }\href {https://doi.org/10.1038/s41467-017-02641-7} {\bibfield  {journal} {\bibinfo
  {journal} {Nat. Commun.}\ }\textbf {\bibinfo {volume} {9}},\ \bibinfo {pages} {102} (\bibinfo {year} {2018})}\BibitemShut {NoStop}%
\bibitem [{\citenamefont {Sakata}\ \emph {et~al.}(2018)\citenamefont {Sakata}, \citenamefont {Lee}, \citenamefont {Morita}, \citenamefont {Johzaki}, \citenamefont {Sawada}, \citenamefont {Iwasa}, \citenamefont {Matsuo}, \citenamefont {Law}, \citenamefont {Yao}, \citenamefont {Hata}, \citenamefont {Sunahara}, \citenamefont {Kojima}, \citenamefont {Abe}, \citenamefont {Kishimoto}, \citenamefont {Syuhada}, \citenamefont {Shiroto}, \citenamefont {Morace}, \citenamefont {Yogo}, \citenamefont {Iwata}, \citenamefont {Nakai}, \citenamefont {Sakagami}, \citenamefont {Ozaki}, \citenamefont {Yamanoi}, \citenamefont {Norimatsu}, \citenamefont {Nakata}, \citenamefont {Tokita}, \citenamefont {Miyanaga}, \citenamefont {Kawanaka}, \citenamefont {Shiraga}, \citenamefont {Mima}, \citenamefont {Nishimura}, \citenamefont {Bailly-Grandvaux}, \citenamefont {Santos}, \citenamefont {Nagatomo}, \citenamefont {Azechi}, \citenamefont {Kodama}, \citenamefont {Arikawa}, \citenamefont {Sentoku},\ and\ \citenamefont
  {Fujioka}}]{Sakata_NC_2018}%
  \BibitemOpen
  \bibfield  {author} {\bibinfo {author} {\bibfnamefont {S.}~\bibnamefont {Sakata}}, \bibinfo {author} {\bibfnamefont {S.}~\bibnamefont {Lee}}, \bibinfo {author} {\bibfnamefont {H.}~\bibnamefont {Morita}}, \bibinfo {author} {\bibfnamefont {T.}~\bibnamefont {Johzaki}}, \bibinfo {author} {\bibfnamefont {H.}~\bibnamefont {Sawada}}, \bibinfo {author} {\bibfnamefont {Y.}~\bibnamefont {Iwasa}}, \bibinfo {author} {\bibfnamefont {K.}~\bibnamefont {Matsuo}}, \bibinfo {author} {\bibfnamefont {K.~F.~F.}\ \bibnamefont {Law}}, \bibinfo {author} {\bibfnamefont {A.}~\bibnamefont {Yao}}, \bibinfo {author} {\bibfnamefont {M.}~\bibnamefont {Hata}}, \bibinfo {author} {\bibfnamefont {A.}~\bibnamefont {Sunahara}}, \bibinfo {author} {\bibfnamefont {S.}~\bibnamefont {Kojima}}, \bibinfo {author} {\bibfnamefont {Y.}~\bibnamefont {Abe}}, \bibinfo {author} {\bibfnamefont {H.}~\bibnamefont {Kishimoto}}, \bibinfo {author} {\bibfnamefont {A.}~\bibnamefont {Syuhada}}, \bibinfo {author} {\bibfnamefont {T.}~\bibnamefont {Shiroto}}, \bibinfo
  {author} {\bibfnamefont {A.}~\bibnamefont {Morace}}, \bibinfo {author} {\bibfnamefont {A.}~\bibnamefont {Yogo}}, \bibinfo {author} {\bibfnamefont {N.}~\bibnamefont {Iwata}}, \bibinfo {author} {\bibfnamefont {M.}~\bibnamefont {Nakai}}, \bibinfo {author} {\bibfnamefont {H.}~\bibnamefont {Sakagami}}, \bibinfo {author} {\bibfnamefont {T.}~\bibnamefont {Ozaki}}, \bibinfo {author} {\bibfnamefont {K.}~\bibnamefont {Yamanoi}}, \bibinfo {author} {\bibfnamefont {T.}~\bibnamefont {Norimatsu}}, \bibinfo {author} {\bibfnamefont {Y.}~\bibnamefont {Nakata}}, \bibinfo {author} {\bibfnamefont {S.}~\bibnamefont {Tokita}}, \bibinfo {author} {\bibfnamefont {N.}~\bibnamefont {Miyanaga}}, \bibinfo {author} {\bibfnamefont {J.}~\bibnamefont {Kawanaka}}, \bibinfo {author} {\bibfnamefont {H.}~\bibnamefont {Shiraga}}, \bibinfo {author} {\bibfnamefont {K.}~\bibnamefont {Mima}}, \bibinfo {author} {\bibfnamefont {H.}~\bibnamefont {Nishimura}}, \bibinfo {author} {\bibfnamefont {M.}~\bibnamefont {Bailly-Grandvaux}}, \bibinfo {author}
  {\bibfnamefont {J.~J.}\ \bibnamefont {Santos}}, \bibinfo {author} {\bibfnamefont {H.}~\bibnamefont {Nagatomo}}, \bibinfo {author} {\bibfnamefont {H.}~\bibnamefont {Azechi}}, \bibinfo {author} {\bibfnamefont {R.}~\bibnamefont {Kodama}}, \bibinfo {author} {\bibfnamefont {Y.}~\bibnamefont {Arikawa}}, \bibinfo {author} {\bibfnamefont {Y.}~\bibnamefont {Sentoku}}, \ and\ \bibinfo {author} {\bibfnamefont {S.}~\bibnamefont {Fujioka}},\ }\href {https://doi.org/10.1038/s41467-018-06173-6} {\bibfield  {journal} {\bibinfo  {journal} {Nat. Commun.}\ }\textbf {\bibinfo {volume} {9}},\ \bibinfo {pages} {3937} (\bibinfo {year} {2018})}\BibitemShut {NoStop}%
\bibitem [{\citenamefont {Matsuo}\ \emph {et~al.}(2021)\citenamefont {Matsuo}, \citenamefont {Sano}, \citenamefont {Nagatomo}, \citenamefont {Somekawa}, \citenamefont {Law}, \citenamefont {Morita}, \citenamefont {Arikawa},\ and\ \citenamefont {Fujioka}}]{Matsuo_PRL_2021}%
  \BibitemOpen
  \bibfield  {author} {\bibinfo {author} {\bibfnamefont {K.}~\bibnamefont {Matsuo}}, \bibinfo {author} {\bibfnamefont {T.}~\bibnamefont {Sano}}, \bibinfo {author} {\bibfnamefont {H.}~\bibnamefont {Nagatomo}}, \bibinfo {author} {\bibfnamefont {T.}~\bibnamefont {Somekawa}}, \bibinfo {author} {\bibfnamefont {K.~F.~F.}\ \bibnamefont {Law}}, \bibinfo {author} {\bibfnamefont {H.}~\bibnamefont {Morita}}, \bibinfo {author} {\bibfnamefont {Y.}~\bibnamefont {Arikawa}}, \ and\ \bibinfo {author} {\bibfnamefont {S.}~\bibnamefont {Fujioka}},\ }\href {\doibase 10.1103/PhysRevLett.127.165001} {\bibfield  {journal} {\bibinfo  {journal} {Phys. Rev. Lett.}\ }\textbf {\bibinfo {volume} {127}},\ \bibinfo {pages} {165001} (\bibinfo {year} {2021})}\BibitemShut {NoStop}%
\bibitem [{\citenamefont {Woolsey}\ \emph {et~al.}(2001)\citenamefont {Woolsey}, \citenamefont {Ali}, \citenamefont {Evans}, \citenamefont {Grundy}, \citenamefont {Pestehe}, \citenamefont {Carolan}, \citenamefont {Conway}, \citenamefont {Dendy}, \citenamefont {Helander}, \citenamefont {McClements}, \citenamefont {Kirk}, \citenamefont {Norreys}, \citenamefont {Notley},\ and\ \citenamefont {Rose}}]{Woolsey_PoP_2001}%
  \BibitemOpen
  \bibfield  {author} {\bibinfo {author} {\bibfnamefont {N.~C.}\ \bibnamefont {Woolsey}}, \bibinfo {author} {\bibfnamefont {Y.~A.}\ \bibnamefont {Ali}}, \bibinfo {author} {\bibfnamefont {R.~G.}\ \bibnamefont {Evans}}, \bibinfo {author} {\bibfnamefont {R.~A.~D.}\ \bibnamefont {Grundy}}, \bibinfo {author} {\bibfnamefont {S.~J.}\ \bibnamefont {Pestehe}}, \bibinfo {author} {\bibfnamefont {P.~G.}\ \bibnamefont {Carolan}}, \bibinfo {author} {\bibfnamefont {N.~J.}\ \bibnamefont {Conway}}, \bibinfo {author} {\bibfnamefont {R.~O.}\ \bibnamefont {Dendy}}, \bibinfo {author} {\bibfnamefont {P.}~\bibnamefont {Helander}}, \bibinfo {author} {\bibfnamefont {K.~G.}\ \bibnamefont {McClements}}, \bibinfo {author} {\bibfnamefont {J.~G.}\ \bibnamefont {Kirk}}, \bibinfo {author} {\bibfnamefont {P.~A.}\ \bibnamefont {Norreys}}, \bibinfo {author} {\bibfnamefont {M.~M.}\ \bibnamefont {Notley}}, \ and\ \bibinfo {author} {\bibfnamefont {S.~J.}\ \bibnamefont {Rose}},\ }\href {\doibase 10.1063/1.1351831} {\bibfield  {journal} {\bibinfo
  {journal} {Physics of Plasmas}\ }\textbf {\bibinfo {volume} {8}},\ \bibinfo {pages} {2439} (\bibinfo {year} {2001})}\BibitemShut {NoStop}%
\bibitem [{\citenamefont {Pisarczyk}\ \emph {et~al.}(2022)\citenamefont {Pisarczyk}, \citenamefont {Renner}, \citenamefont {Dudzak}, \citenamefont {Chodukowski}, \citenamefont {Rusiniak}, \citenamefont {Domanski}, \citenamefont {Badziak}, \citenamefont {Dostal}, \citenamefont {Krupka}, \citenamefont {Singh}, \citenamefont {Klir}, \citenamefont {Ehret}, \citenamefont {Gajdos}, \citenamefont {Zaras-Szydłowska}, \citenamefont {Rosinski}, \citenamefont {Tchórz}, \citenamefont {Szymanski}, \citenamefont {Krasa}, \citenamefont {Burian}, \citenamefont {Pfeifer}, \citenamefont {Cikhardt}, \citenamefont {Jelinek}, \citenamefont {Kocourkova}, \citenamefont {Batani}, \citenamefont {Batani}, \citenamefont {Santos}, \citenamefont {Vlachos}, \citenamefont {Ospina-Bohórquez}, \citenamefont {Volpe}, \citenamefont {Borodziuk}, \citenamefont {Krus},\ and\ \citenamefont {Juha}}]{Pisarczyk_PPCF_2022}%
  \BibitemOpen
  \bibfield  {author} {\bibinfo {author} {\bibfnamefont {T.}~\bibnamefont {Pisarczyk}}, \bibinfo {author} {\bibfnamefont {O.}~\bibnamefont {Renner}}, \bibinfo {author} {\bibfnamefont {R.}~\bibnamefont {Dudzak}}, \bibinfo {author} {\bibfnamefont {T.}~\bibnamefont {Chodukowski}}, \bibinfo {author} {\bibfnamefont {Z.}~\bibnamefont {Rusiniak}}, \bibinfo {author} {\bibfnamefont {J.}~\bibnamefont {Domanski}}, \bibinfo {author} {\bibfnamefont {J.}~\bibnamefont {Badziak}}, \bibinfo {author} {\bibfnamefont {J.}~\bibnamefont {Dostal}}, \bibinfo {author} {\bibfnamefont {M.}~\bibnamefont {Krupka}}, \bibinfo {author} {\bibfnamefont {S.}~\bibnamefont {Singh}}, \bibinfo {author} {\bibfnamefont {D.}~\bibnamefont {Klir}}, \bibinfo {author} {\bibfnamefont {M.}~\bibnamefont {Ehret}}, \bibinfo {author} {\bibfnamefont {P.}~\bibnamefont {Gajdos}}, \bibinfo {author} {\bibfnamefont {A.}~\bibnamefont {Zaras-Szydłowska}}, \bibinfo {author} {\bibfnamefont {M.}~\bibnamefont {Rosinski}}, \bibinfo {author} {\bibfnamefont
  {P.}~\bibnamefont {Tchórz}}, \bibinfo {author} {\bibfnamefont {M.}~\bibnamefont {Szymanski}}, \bibinfo {author} {\bibfnamefont {J.}~\bibnamefont {Krasa}}, \bibinfo {author} {\bibfnamefont {T.}~\bibnamefont {Burian}}, \bibinfo {author} {\bibfnamefont {M.}~\bibnamefont {Pfeifer}}, \bibinfo {author} {\bibfnamefont {J.}~\bibnamefont {Cikhardt}}, \bibinfo {author} {\bibfnamefont {S.}~\bibnamefont {Jelinek}}, \bibinfo {author} {\bibfnamefont {G.}~\bibnamefont {Kocourkova}}, \bibinfo {author} {\bibfnamefont {D.}~\bibnamefont {Batani}}, \bibinfo {author} {\bibfnamefont {K.}~\bibnamefont {Batani}}, \bibinfo {author} {\bibfnamefont {J.}~\bibnamefont {Santos}}, \bibinfo {author} {\bibfnamefont {C.}~\bibnamefont {Vlachos}}, \bibinfo {author} {\bibfnamefont {V.}~\bibnamefont {Ospina-Bohórquez}}, \bibinfo {author} {\bibfnamefont {L.}~\bibnamefont {Volpe}}, \bibinfo {author} {\bibfnamefont {S.}~\bibnamefont {Borodziuk}}, \bibinfo {author} {\bibfnamefont {M.}~\bibnamefont {Krus}}, \ and\ \bibinfo {author} {\bibfnamefont
  {L.}~\bibnamefont {Juha}},\ }\href {\doibase 10.1088/1361-6587/ac95c4} {\bibfield  {journal} {\bibinfo  {journal} {Plasma Physics and Controlled Fusion}\ }\textbf {\bibinfo {volume} {64}},\ \bibinfo {pages} {115012} (\bibinfo {year} {2022})}\BibitemShut {NoStop}%
\bibitem [{\citenamefont {Chien}\ \emph {et~al.}(2019)\citenamefont {Chien}, \citenamefont {Gao}, \citenamefont {Ji}, \citenamefont {Yuan}, \citenamefont {Blackman}, \citenamefont {Chen}, \citenamefont {Efthimion}, \citenamefont {Fiksel}, \citenamefont {Froula}, \citenamefont {Hill}, \citenamefont {Huang}, \citenamefont {Lu}, \citenamefont {Moody},\ and\ \citenamefont {Nilson}}]{chien2019}%
  \BibitemOpen
  \bibfield  {author} {\bibinfo {author} {\bibfnamefont {A.}~\bibnamefont {Chien}}, \bibinfo {author} {\bibfnamefont {L.}~\bibnamefont {Gao}}, \bibinfo {author} {\bibfnamefont {H.}~\bibnamefont {Ji}}, \bibinfo {author} {\bibfnamefont {X.}~\bibnamefont {Yuan}}, \bibinfo {author} {\bibfnamefont {E.~G.}\ \bibnamefont {Blackman}}, \bibinfo {author} {\bibfnamefont {H.}~\bibnamefont {Chen}}, \bibinfo {author} {\bibfnamefont {P.~C.}\ \bibnamefont {Efthimion}}, \bibinfo {author} {\bibfnamefont {G.}~\bibnamefont {Fiksel}}, \bibinfo {author} {\bibfnamefont {D.~H.}\ \bibnamefont {Froula}}, \bibinfo {author} {\bibfnamefont {K.~W.}\ \bibnamefont {Hill}}, \bibinfo {author} {\bibfnamefont {K.}~\bibnamefont {Huang}}, \bibinfo {author} {\bibfnamefont {Q.}~\bibnamefont {Lu}}, \bibinfo {author} {\bibfnamefont {J.~D.}\ \bibnamefont {Moody}}, \ and\ \bibinfo {author} {\bibfnamefont {P.~M.}\ \bibnamefont {Nilson}},\ }\href {\doibase 10.1063/1.5095960} {\bibfield  {journal} {\bibinfo  {journal} {Physics of Plasmas}\ }\textbf
  {\bibinfo {volume} {26}},\ \bibinfo {pages} {062113} (\bibinfo {year} {2019})}\BibitemShut {NoStop}%
\bibitem [{\citenamefont {Chien}\ \emph {et~al.}(2023)\citenamefont {Chien}, \citenamefont {Gao}, \citenamefont {Zhang}, \citenamefont {Ji}, \citenamefont {Blackman}, \citenamefont {Daughton}, \citenamefont {Stanier}, \citenamefont {Le}, \citenamefont {Guo}, \citenamefont {Follett}, \citenamefont {Chen}, \citenamefont {Fiksel}, \citenamefont {Bleotu}, \citenamefont {Cauble}, \citenamefont {Chen}, \citenamefont {Fazzini}, \citenamefont {Flippo}, \citenamefont {French}, \citenamefont {Froula}, \citenamefont {Fuchs}, \citenamefont {Fujioka}, \citenamefont {Hill}, \citenamefont {Klein}, \citenamefont {Kuranz}, \citenamefont {Nilson}, \citenamefont {Rasmus},\ and\ \citenamefont {Takizawa}}]{chien23}%
  \BibitemOpen
  \bibfield  {author} {\bibinfo {author} {\bibfnamefont {A.}~\bibnamefont {Chien}}, \bibinfo {author} {\bibfnamefont {L.}~\bibnamefont {Gao}}, \bibinfo {author} {\bibfnamefont {S.}~\bibnamefont {Zhang}}, \bibinfo {author} {\bibfnamefont {H.}~\bibnamefont {Ji}}, \bibinfo {author} {\bibfnamefont {E.~G.}\ \bibnamefont {Blackman}}, \bibinfo {author} {\bibfnamefont {W.}~\bibnamefont {Daughton}}, \bibinfo {author} {\bibfnamefont {A.}~\bibnamefont {Stanier}}, \bibinfo {author} {\bibfnamefont {A.}~\bibnamefont {Le}}, \bibinfo {author} {\bibfnamefont {F.}~\bibnamefont {Guo}}, \bibinfo {author} {\bibfnamefont {R.}~\bibnamefont {Follett}}, \bibinfo {author} {\bibfnamefont {H.}~\bibnamefont {Chen}}, \bibinfo {author} {\bibfnamefont {G.}~\bibnamefont {Fiksel}}, \bibinfo {author} {\bibfnamefont {G.}~\bibnamefont {Bleotu}}, \bibinfo {author} {\bibfnamefont {R.~C.}\ \bibnamefont {Cauble}}, \bibinfo {author} {\bibfnamefont {S.~N.}\ \bibnamefont {Chen}}, \bibinfo {author} {\bibfnamefont {A.}~\bibnamefont {Fazzini}}, \bibinfo
  {author} {\bibfnamefont {K.}~\bibnamefont {Flippo}}, \bibinfo {author} {\bibfnamefont {O.}~\bibnamefont {French}}, \bibinfo {author} {\bibfnamefont {D.~H.}\ \bibnamefont {Froula}}, \bibinfo {author} {\bibfnamefont {J.}~\bibnamefont {Fuchs}}, \bibinfo {author} {\bibfnamefont {S.}~\bibnamefont {Fujioka}}, \bibinfo {author} {\bibfnamefont {K.}~\bibnamefont {Hill}}, \bibinfo {author} {\bibfnamefont {S.}~\bibnamefont {Klein}}, \bibinfo {author} {\bibfnamefont {C.}~\bibnamefont {Kuranz}}, \bibinfo {author} {\bibfnamefont {P.}~\bibnamefont {Nilson}}, \bibinfo {author} {\bibfnamefont {A.}~\bibnamefont {Rasmus}}, \ and\ \bibinfo {author} {\bibfnamefont {R.}~\bibnamefont {Takizawa}},\ }\href {https://doi.org/10.1038/s41567-022-01839-x} {\bibfield  {journal} {\bibinfo  {journal} {Nature Physics}\ }\textbf {\bibinfo {volume} {19}},\ \bibinfo {pages} {254} (\bibinfo {year} {2023})}\BibitemShut {NoStop}%
\bibitem [{\citenamefont {{Zhang}}\ \emph {et~al.}(2023)\citenamefont {{Zhang}}, \citenamefont {{Chien}}, \citenamefont {{Gao}}, \citenamefont {{Ji}}, \citenamefont {{Blackman}}, \citenamefont {{Follett}}, \citenamefont {{Froula}}, \citenamefont {{Katz}}, \citenamefont {{Li}}, \citenamefont {{Birkel}}, \citenamefont {{Petrasso}}, \citenamefont {{Moody}},\ and\ \citenamefont {{Chen}}}]{zhang23}%
  \BibitemOpen
  \bibfield  {author} {\bibinfo {author} {\bibfnamefont {S.}~\bibnamefont {{Zhang}}}, \bibinfo {author} {\bibfnamefont {A.}~\bibnamefont {{Chien}}}, \bibinfo {author} {\bibfnamefont {L.}~\bibnamefont {{Gao}}}, \bibinfo {author} {\bibfnamefont {H.}~\bibnamefont {{Ji}}}, \bibinfo {author} {\bibfnamefont {E.~G.}\ \bibnamefont {{Blackman}}}, \bibinfo {author} {\bibfnamefont {R.}~\bibnamefont {{Follett}}}, \bibinfo {author} {\bibfnamefont {D.~H.}\ \bibnamefont {{Froula}}}, \bibinfo {author} {\bibfnamefont {J.}~\bibnamefont {{Katz}}}, \bibinfo {author} {\bibfnamefont {C.}~\bibnamefont {{Li}}}, \bibinfo {author} {\bibfnamefont {A.}~\bibnamefont {{Birkel}}}, \bibinfo {author} {\bibfnamefont {R.}~\bibnamefont {{Petrasso}}}, \bibinfo {author} {\bibfnamefont {J.}~\bibnamefont {{Moody}}}, \ and\ \bibinfo {author} {\bibfnamefont {H.}~\bibnamefont {{Chen}}},\ }\href {\doibase 10.1038/s41567-023-01972-1} {\bibfield  {journal} {\bibinfo  {journal} {Nature Physics}\ }\textbf {\bibinfo {volume} {19}},\ \bibinfo {pages} {909}
  (\bibinfo {year} {2023})}\BibitemShut {NoStop}%
\bibitem [{\citenamefont {Ji}\ \emph {et~al.}(2024)\citenamefont {Ji}, \citenamefont {Gao}, \citenamefont {Pomraning}, \citenamefont {Sakai}, \citenamefont {Guo}, \citenamefont {Li}, \citenamefont {Stanier}, \citenamefont {Milder}, \citenamefont {Follett}, \citenamefont {Fiksel}, \citenamefont {Blackman}, \citenamefont {Chien},\ and\ \citenamefont {Zhang}}]{ji_PoP_2024}%
  \BibitemOpen
  \bibfield  {author} {\bibinfo {author} {\bibfnamefont {H.}~\bibnamefont {Ji}}, \bibinfo {author} {\bibfnamefont {L.}~\bibnamefont {Gao}}, \bibinfo {author} {\bibfnamefont {G.}~\bibnamefont {Pomraning}}, \bibinfo {author} {\bibfnamefont {K.}~\bibnamefont {Sakai}}, \bibinfo {author} {\bibfnamefont {F.}~\bibnamefont {Guo}}, \bibinfo {author} {\bibfnamefont {X.}~\bibnamefont {Li}}, \bibinfo {author} {\bibfnamefont {A.}~\bibnamefont {Stanier}}, \bibinfo {author} {\bibfnamefont {A.}~\bibnamefont {Milder}}, \bibinfo {author} {\bibfnamefont {R.~K.}\ \bibnamefont {Follett}}, \bibinfo {author} {\bibfnamefont {G.}~\bibnamefont {Fiksel}}, \bibinfo {author} {\bibfnamefont {E.~G.}\ \bibnamefont {Blackman}}, \bibinfo {author} {\bibfnamefont {A.}~\bibnamefont {Chien}}, \ and\ \bibinfo {author} {\bibfnamefont {S.}~\bibnamefont {Zhang}},\ }\href {\doibase 10.1063/5.0223922} {\bibfield  {journal} {\bibinfo  {journal} {Physics of Plasmas}\ }\textbf {\bibinfo {volume} {31}},\ \bibinfo {pages} {102112} (\bibinfo {year}
  {2024})}\BibitemShut {NoStop}%
\bibitem [{\citenamefont {Perkins}\ \emph {et~al.}(2013)\citenamefont {Perkins}, \citenamefont {Logan}, \citenamefont {Zimmerman},\ and\ \citenamefont {Werner}}]{Perkins_PoP_2013}%
  \BibitemOpen
  \bibfield  {author} {\bibinfo {author} {\bibfnamefont {L.~J.}\ \bibnamefont {Perkins}}, \bibinfo {author} {\bibfnamefont {B.~G.}\ \bibnamefont {Logan}}, \bibinfo {author} {\bibfnamefont {G.~B.}\ \bibnamefont {Zimmerman}}, \ and\ \bibinfo {author} {\bibfnamefont {C.~J.}\ \bibnamefont {Werner}},\ }\href {\doibase 10.1063/1.4816813} {\bibfield  {journal} {\bibinfo  {journal} {Physics of Plasmas}\ }\textbf {\bibinfo {volume} {20}},\ \bibinfo {pages} {072708} (\bibinfo {year} {2013})}\BibitemShut {NoStop}%
\bibitem [{\citenamefont {Beg}\ \emph {et~al.}(1997)\citenamefont {Beg}, \citenamefont {Bell}, \citenamefont {Dangor}, \citenamefont {Danson}, \citenamefont {Fews}, \citenamefont {Glinsky}, \citenamefont {Hammel}, \citenamefont {Lee}, \citenamefont {Norreys},\ and\ \citenamefont {Tatarakis}}]{Beg_PoP_1997}%
  \BibitemOpen
  \bibfield  {author} {\bibinfo {author} {\bibfnamefont {F.~N.}\ \bibnamefont {Beg}}, \bibinfo {author} {\bibfnamefont {A.~R.}\ \bibnamefont {Bell}}, \bibinfo {author} {\bibfnamefont {A.~E.}\ \bibnamefont {Dangor}}, \bibinfo {author} {\bibfnamefont {C.~N.}\ \bibnamefont {Danson}}, \bibinfo {author} {\bibfnamefont {A.~P.}\ \bibnamefont {Fews}}, \bibinfo {author} {\bibfnamefont {M.~E.}\ \bibnamefont {Glinsky}}, \bibinfo {author} {\bibfnamefont {B.~A.}\ \bibnamefont {Hammel}}, \bibinfo {author} {\bibfnamefont {P.}~\bibnamefont {Lee}}, \bibinfo {author} {\bibfnamefont {P.~A.}\ \bibnamefont {Norreys}}, \ and\ \bibinfo {author} {\bibfnamefont {M.}~\bibnamefont {Tatarakis}},\ }\href {\doibase 10.1063/1.872103} {\bibfield  {journal} {\bibinfo  {journal} {Physics of Plasmas}\ }\textbf {\bibinfo {volume} {4}},\ \bibinfo {pages} {447} (\bibinfo {year} {1997})}\BibitemShut {NoStop}%
\bibitem [{\citenamefont {Li}\ \emph {et~al.}(2004)\citenamefont {Li}, \citenamefont {Zhang}, \citenamefont {Sheng}, \citenamefont {Zheng}, \citenamefont {Chen}, \citenamefont {Kodama}, \citenamefont {Matsuoka}, \citenamefont {Tampo}, \citenamefont {Tanaka}, \citenamefont {Tsutsumi},\ and\ \citenamefont {Yabuuchi}}]{YTLi_PRE_2004}%
  \BibitemOpen
  \bibfield  {author} {\bibinfo {author} {\bibfnamefont {Y.~T.}\ \bibnamefont {Li}}, \bibinfo {author} {\bibfnamefont {J.}~\bibnamefont {Zhang}}, \bibinfo {author} {\bibfnamefont {Z.~M.}\ \bibnamefont {Sheng}}, \bibinfo {author} {\bibfnamefont {J.}~\bibnamefont {Zheng}}, \bibinfo {author} {\bibfnamefont {Z.~L.}\ \bibnamefont {Chen}}, \bibinfo {author} {\bibfnamefont {R.}~\bibnamefont {Kodama}}, \bibinfo {author} {\bibfnamefont {T.}~\bibnamefont {Matsuoka}}, \bibinfo {author} {\bibfnamefont {M.}~\bibnamefont {Tampo}}, \bibinfo {author} {\bibfnamefont {K.~A.}\ \bibnamefont {Tanaka}}, \bibinfo {author} {\bibfnamefont {T.}~\bibnamefont {Tsutsumi}}, \ and\ \bibinfo {author} {\bibfnamefont {T.}~\bibnamefont {Yabuuchi}},\ }\href {\doibase 10.1103/PhysRevE.69.036405} {\bibfield  {journal} {\bibinfo  {journal} {Phys. Rev. E}\ }\textbf {\bibinfo {volume} {69}},\ \bibinfo {pages} {036405} (\bibinfo {year} {2004})}\BibitemShut {NoStop}%
\bibitem [{\citenamefont {Wang}\ \emph {et~al.}(2023)\citenamefont {Wang}, \citenamefont {Shan}, \citenamefont {Zhang}, \citenamefont {Yuan}, \citenamefont {Liu}, \citenamefont {Tian}, \citenamefont {Yang}, \citenamefont {Lu}, \citenamefont {Qi}, \citenamefont {Deng}, \citenamefont {Zhou}, \citenamefont {Xie}, \citenamefont {Wang}, \citenamefont {Mu}, \citenamefont {Zhou}, \citenamefont {Cai}, \citenamefont {Zhu},\ and\ \citenamefont {Gu}}]{Wang_PoP_2023}%
  \BibitemOpen
  \bibfield  {author} {\bibinfo {author} {\bibfnamefont {W.}~\bibnamefont {Wang}}, \bibinfo {author} {\bibfnamefont {L.}~\bibnamefont {Shan}}, \bibinfo {author} {\bibfnamefont {F.}~\bibnamefont {Zhang}}, \bibinfo {author} {\bibfnamefont {Z.}~\bibnamefont {Yuan}}, \bibinfo {author} {\bibfnamefont {D.}~\bibnamefont {Liu}}, \bibinfo {author} {\bibfnamefont {C.}~\bibnamefont {Tian}}, \bibinfo {author} {\bibfnamefont {L.}~\bibnamefont {Yang}}, \bibinfo {author} {\bibfnamefont {F.}~\bibnamefont {Lu}}, \bibinfo {author} {\bibfnamefont {W.}~\bibnamefont {Qi}}, \bibinfo {author} {\bibfnamefont {Z.}~\bibnamefont {Deng}}, \bibinfo {author} {\bibfnamefont {K.}~\bibnamefont {Zhou}}, \bibinfo {author} {\bibfnamefont {N.}~\bibnamefont {Xie}}, \bibinfo {author} {\bibfnamefont {X.}~\bibnamefont {Wang}}, \bibinfo {author} {\bibfnamefont {J.}~\bibnamefont {Mu}}, \bibinfo {author} {\bibfnamefont {W.}~\bibnamefont {Zhou}}, \bibinfo {author} {\bibfnamefont {H.}~\bibnamefont {Cai}}, \bibinfo {author} {\bibfnamefont
  {S.}~\bibnamefont {Zhu}}, \ and\ \bibinfo {author} {\bibfnamefont {Y.}~\bibnamefont {Gu}},\ }\href {\doibase 10.1063/5.0120697} {\bibfield  {journal} {\bibinfo  {journal} {Physics of Plasmas}\ }\textbf {\bibinfo {volume} {30}},\ \bibinfo {pages} {072703} (\bibinfo {year} {2023})}\BibitemShut {NoStop}%
\bibitem [{\citenamefont {Gao}\ \emph {et~al.}(2012)\citenamefont {Gao}, \citenamefont {Nilson}, \citenamefont {Igumenschev}, \citenamefont {Hu}, \citenamefont {Davies}, \citenamefont {Stoeckl}, \citenamefont {Haines}, \citenamefont {Froula}, \citenamefont {Betti},\ and\ \citenamefont {Meyerhofer}}]{Gao2012}%
  \BibitemOpen
  \bibfield  {author} {\bibinfo {author} {\bibfnamefont {L.}~\bibnamefont {Gao}}, \bibinfo {author} {\bibfnamefont {P.~M.}\ \bibnamefont {Nilson}}, \bibinfo {author} {\bibfnamefont {I.~V.}\ \bibnamefont {Igumenschev}}, \bibinfo {author} {\bibfnamefont {S.~X.}\ \bibnamefont {Hu}}, \bibinfo {author} {\bibfnamefont {J.~R.}\ \bibnamefont {Davies}}, \bibinfo {author} {\bibfnamefont {C.}~\bibnamefont {Stoeckl}}, \bibinfo {author} {\bibfnamefont {M.~G.}\ \bibnamefont {Haines}}, \bibinfo {author} {\bibfnamefont {D.~H.}\ \bibnamefont {Froula}}, \bibinfo {author} {\bibfnamefont {R.}~\bibnamefont {Betti}}, \ and\ \bibinfo {author} {\bibfnamefont {D.~D.}\ \bibnamefont {Meyerhofer}},\ }\href {\doibase 10.1103/PhysRevLett.109.115001} {\bibfield  {journal} {\bibinfo  {journal} {Phys. Rev. Lett.}\ }\textbf {\bibinfo {volume} {109}},\ \bibinfo {pages} {115001} (\bibinfo {year} {2012})}\BibitemShut {NoStop}%
\bibitem [{\citenamefont {Gao}\ \emph {et~al.}(2013)\citenamefont {Gao}, \citenamefont {Nilson}, \citenamefont {Igumenschev}, \citenamefont {Fiksel}, \citenamefont {Yan}, \citenamefont {Davies}, \citenamefont {Martinez}, \citenamefont {Smalyuk}, \citenamefont {Haines}, \citenamefont {Blackman}, \citenamefont {Froula}, \citenamefont {Betti},\ and\ \citenamefont {Meyerhofer}}]{Gao2013}%
  \BibitemOpen
  \bibfield  {author} {\bibinfo {author} {\bibfnamefont {L.}~\bibnamefont {Gao}}, \bibinfo {author} {\bibfnamefont {P.~M.}\ \bibnamefont {Nilson}}, \bibinfo {author} {\bibfnamefont {I.~V.}\ \bibnamefont {Igumenschev}}, \bibinfo {author} {\bibfnamefont {G.}~\bibnamefont {Fiksel}}, \bibinfo {author} {\bibfnamefont {R.}~\bibnamefont {Yan}}, \bibinfo {author} {\bibfnamefont {J.~R.}\ \bibnamefont {Davies}}, \bibinfo {author} {\bibfnamefont {D.}~\bibnamefont {Martinez}}, \bibinfo {author} {\bibfnamefont {V.}~\bibnamefont {Smalyuk}}, \bibinfo {author} {\bibfnamefont {M.~G.}\ \bibnamefont {Haines}}, \bibinfo {author} {\bibfnamefont {E.~G.}\ \bibnamefont {Blackman}}, \bibinfo {author} {\bibfnamefont {D.~H.}\ \bibnamefont {Froula}}, \bibinfo {author} {\bibfnamefont {R.}~\bibnamefont {Betti}}, \ and\ \bibinfo {author} {\bibfnamefont {D.~D.}\ \bibnamefont {Meyerhofer}},\ }\href {\doibase 10.1103/PhysRevLett.110.185003} {\bibfield  {journal} {\bibinfo  {journal} {Phys. Rev. Lett.}\ }\textbf {\bibinfo {volume} {110}},\
  \bibinfo {pages} {185003} (\bibinfo {year} {2013})}\BibitemShut {NoStop}%
\bibitem [{\citenamefont {Gao}\ \emph {et~al.}(2015)\citenamefont {Gao}, \citenamefont {Nilson}, \citenamefont {Igumenshchev}, \citenamefont {Haines}, \citenamefont {Froula}, \citenamefont {Betti},\ and\ \citenamefont {Meyerhofer}}]{Gao2015}%
  \BibitemOpen
  \bibfield  {author} {\bibinfo {author} {\bibfnamefont {L.}~\bibnamefont {Gao}}, \bibinfo {author} {\bibfnamefont {P.~M.}\ \bibnamefont {Nilson}}, \bibinfo {author} {\bibfnamefont {I.~V.}\ \bibnamefont {Igumenshchev}}, \bibinfo {author} {\bibfnamefont {M.~G.}\ \bibnamefont {Haines}}, \bibinfo {author} {\bibfnamefont {D.~H.}\ \bibnamefont {Froula}}, \bibinfo {author} {\bibfnamefont {R.}~\bibnamefont {Betti}}, \ and\ \bibinfo {author} {\bibfnamefont {D.~D.}\ \bibnamefont {Meyerhofer}},\ }\href {\doibase 10.1103/PhysRevLett.114.215003} {\bibfield  {journal} {\bibinfo  {journal} {Phys. Rev. Lett.}\ }\textbf {\bibinfo {volume} {114}},\ \bibinfo {pages} {215003} (\bibinfo {year} {2015})}\BibitemShut {NoStop}%
\bibitem [{\citenamefont {Waxer}\ \emph {et~al.}(2005)\citenamefont {Waxer}, \citenamefont {D.~N.~Maywar}, \citenamefont {B.~E.~Kruschwitz}, \citenamefont {D.~D.~Meyerhofer},\ and\ \citenamefont {Zuegel}}]{Waxer2005}%
  \BibitemOpen
  \bibfield  {author} {\bibinfo {author} {\bibfnamefont {L.~J.}\ \bibnamefont {Waxer}}, \bibinfo {author} {\bibfnamefont {T.~J.~K.}\ \bibnamefont {D.~N.~Maywar}, \bibfnamefont {J.~H.~Kelly}}, \bibinfo {author} {\bibfnamefont {R.~L.~M.}\ \bibnamefont {B.~E.~Kruschwitz}, \bibfnamefont {S.~J.~Loucks}}, \bibinfo {author} {\bibfnamefont {C.~S.}\ \bibnamefont {D.~D.~Meyerhofer}, \bibfnamefont {S.~F. B.~Morse}}, \ and\ \bibinfo {author} {\bibfnamefont {J.~D.}\ \bibnamefont {Zuegel}},\ }\href {\doibase 10.1364/OPN.16.7.000030} {\bibfield  {journal} {\bibinfo  {journal} {Opt. Photon. News}\ }\textbf {\bibinfo {volume} {16}},\ \bibinfo {pages} {30} (\bibinfo {year} {2005})}\BibitemShut {NoStop}%
\bibitem [{\citenamefont {Wilks}\ \emph {et~al.}(2001)\citenamefont {Wilks}, \citenamefont {Langdon}, \citenamefont {Cowan}, \citenamefont {Roth}, \citenamefont {Singh}, \citenamefont {Hatchett}, \citenamefont {Key}, \citenamefont {Pennington}, \citenamefont {MacKinnon},\ and\ \citenamefont {Snavely}}]{wilks2001}%
  \BibitemOpen
  \bibfield  {author} {\bibinfo {author} {\bibfnamefont {S.~C.}\ \bibnamefont {Wilks}}, \bibinfo {author} {\bibfnamefont {A.~B.}\ \bibnamefont {Langdon}}, \bibinfo {author} {\bibfnamefont {T.~E.}\ \bibnamefont {Cowan}}, \bibinfo {author} {\bibfnamefont {M.}~\bibnamefont {Roth}}, \bibinfo {author} {\bibfnamefont {M.}~\bibnamefont {Singh}}, \bibinfo {author} {\bibfnamefont {S.}~\bibnamefont {Hatchett}}, \bibinfo {author} {\bibfnamefont {M.~H.}\ \bibnamefont {Key}}, \bibinfo {author} {\bibfnamefont {D.}~\bibnamefont {Pennington}}, \bibinfo {author} {\bibfnamefont {A.}~\bibnamefont {MacKinnon}}, \ and\ \bibinfo {author} {\bibfnamefont {R.~A.}\ \bibnamefont {Snavely}},\ }\href {\doibase 10.1063/1.1333697} {\bibfield  {journal} {\bibinfo  {journal} {Phys. Plasmas}\ }\textbf {\bibinfo {volume} {8}},\ \bibinfo {pages} {542} (\bibinfo {year} {2001})}\BibitemShut {NoStop}%
\bibitem [{\citenamefont {Zylstra}\ \emph {et~al.}(2012)\citenamefont {Zylstra}, \citenamefont {Li}, \citenamefont {Rinderknecht}, \citenamefont {Seguin}, \citenamefont {Petrasso}, \citenamefont {Stoeckl}, \citenamefont {Meyerhofer}, \citenamefont {Nilson}, \citenamefont {Sangster}, \citenamefont {Pape}, \citenamefont {Mackinnon},\ and\ \citenamefont {Patel}}]{zylstra:013511}%
  \BibitemOpen
  \bibfield  {author} {\bibinfo {author} {\bibfnamefont {A.~B.}\ \bibnamefont {Zylstra}}, \bibinfo {author} {\bibfnamefont {C.~K.}\ \bibnamefont {Li}}, \bibinfo {author} {\bibfnamefont {H.~G.}\ \bibnamefont {Rinderknecht}}, \bibinfo {author} {\bibfnamefont {F.~H.}\ \bibnamefont {Seguin}}, \bibinfo {author} {\bibfnamefont {R.~D.}\ \bibnamefont {Petrasso}}, \bibinfo {author} {\bibfnamefont {C.}~\bibnamefont {Stoeckl}}, \bibinfo {author} {\bibfnamefont {D.~D.}\ \bibnamefont {Meyerhofer}}, \bibinfo {author} {\bibfnamefont {P.}~\bibnamefont {Nilson}}, \bibinfo {author} {\bibfnamefont {T.~C.}\ \bibnamefont {Sangster}}, \bibinfo {author} {\bibfnamefont {S.~L.}\ \bibnamefont {Pape}}, \bibinfo {author} {\bibfnamefont {A.}~\bibnamefont {Mackinnon}}, \ and\ \bibinfo {author} {\bibfnamefont {P.}~\bibnamefont {Patel}},\ }\href {\doibase 10.1063/1.3680110} {\bibfield  {journal} {\bibinfo  {journal} {Rev. Sci. Instrum.}\ }\textbf {\bibinfo {volume} {83}},\ \bibinfo {eid} {013511} (\bibinfo {year} {2012})}\BibitemShut
  {NoStop}%
\bibitem [{\citenamefont {{Gao}}(2014)}]{gao14thesis}%
  \BibitemOpen
  \bibfield  {author} {\bibinfo {author} {\bibfnamefont {L.}~\bibnamefont {{Gao}}},\ }\emph {\bibinfo {title} {{Measurements of Magneto-Hydrodynamic Effects in Ablatively-Driven High Energy Density Systems}}},\ \href@noop {} {Ph.D. thesis},\ \bibinfo  {school} {University of Rochester} (\bibinfo {year} {2014})\BibitemShut {NoStop}%
\bibitem [{\citenamefont {Borghesi}\ \emph {et~al.}(2001)\citenamefont {Borghesi}, \citenamefont {Schiavi}, \citenamefont {Campbell}, \citenamefont {Haines}, \citenamefont {Willi}, \citenamefont {MacKinnon}, \citenamefont {Gizzi}, \citenamefont {Galimberti}, \citenamefont {Clarke},\ and\ \citenamefont {Ruhl}}]{Borghesi_ppcf_ProtonImaging}%
  \BibitemOpen
  \bibfield  {author} {\bibinfo {author} {\bibfnamefont {M.}~\bibnamefont {Borghesi}}, \bibinfo {author} {\bibfnamefont {A.}~\bibnamefont {Schiavi}}, \bibinfo {author} {\bibfnamefont {D.~H.}\ \bibnamefont {Campbell}}, \bibinfo {author} {\bibfnamefont {M.~G.}\ \bibnamefont {Haines}}, \bibinfo {author} {\bibfnamefont {O.}~\bibnamefont {Willi}}, \bibinfo {author} {\bibfnamefont {A.~J.}\ \bibnamefont {MacKinnon}}, \bibinfo {author} {\bibfnamefont {L.~A.}\ \bibnamefont {Gizzi}}, \bibinfo {author} {\bibfnamefont {M.}~\bibnamefont {Galimberti}}, \bibinfo {author} {\bibfnamefont {R.~J.}\ \bibnamefont {Clarke}}, \ and\ \bibinfo {author} {\bibfnamefont {H.}~\bibnamefont {Ruhl}},\ }\href {\doibase 10.1088/0741-3335/43/12A/320} {\bibfield  {journal} {\bibinfo  {journal} {Plasma Phys. Controlled Fusion}\ }\textbf {\bibinfo {volume} {43}},\ \bibinfo {pages} {A267} (\bibinfo {year} {2001})}\BibitemShut {NoStop}%
\bibitem [{\citenamefont {Peebles}\ \emph {et~al.}(2022)\citenamefont {Peebles}, \citenamefont {Davies}, \citenamefont {Barnak}, \citenamefont {Garcia-Rubio}, \citenamefont {Heuer}, \citenamefont {Brent}, \citenamefont {Spielman},\ and\ \citenamefont {Betti}}]{Peebles_PoP_2022}%
  \BibitemOpen
  \bibfield  {author} {\bibinfo {author} {\bibfnamefont {J.~L.}\ \bibnamefont {Peebles}}, \bibinfo {author} {\bibfnamefont {J.~R.}\ \bibnamefont {Davies}}, \bibinfo {author} {\bibfnamefont {D.~H.}\ \bibnamefont {Barnak}}, \bibinfo {author} {\bibfnamefont {F.}~\bibnamefont {Garcia-Rubio}}, \bibinfo {author} {\bibfnamefont {P.~V.}\ \bibnamefont {Heuer}}, \bibinfo {author} {\bibfnamefont {G.}~\bibnamefont {Brent}}, \bibinfo {author} {\bibfnamefont {R.}~\bibnamefont {Spielman}}, \ and\ \bibinfo {author} {\bibfnamefont {R.}~\bibnamefont {Betti}},\ }\href {\doibase 10.1063/5.0096784} {\bibfield  {journal} {\bibinfo  {journal} {Physics of Plasmas}\ }\textbf {\bibinfo {volume} {29}},\ \bibinfo {pages} {080501} (\bibinfo {year} {2022})}\BibitemShut {NoStop}%
\bibitem [{\citenamefont {Community}\ \emph {et~al.}(2024)\citenamefont {Community}, \citenamefont {Murphy}, \citenamefont {Everson}, \citenamefont {Stańczak-Marikin}, \citenamefont {Heuer}, \citenamefont {Kozlowski}, \citenamefont {Addison}, \citenamefont {Ahamed}, \citenamefont {Arran}, \citenamefont {Bagherianlemraski}, \citenamefont {Beckers}, \citenamefont {Bedmutha}, \citenamefont {Bergeron}, \citenamefont {Bessi}, \citenamefont {Britten}, \citenamefont {Brown}, \citenamefont {Bryant}, \citenamefont {Carroll}, \citenamefont {Cartagena-Sanchez}, \citenamefont {Chambers}, \citenamefont {Chattopadhyay}, \citenamefont {Choubey}, \citenamefont {Choudhary}, \citenamefont {Clauss}, \citenamefont {Deal}, \citenamefont {Decristoforo}, \citenamefont {Diaz~Riega}, \citenamefont {Dover}, \citenamefont {Drozdov}, \citenamefont {Du}, \citenamefont {Einhorn}, \citenamefont {Fan}, \citenamefont {Farid}, \citenamefont {Fischer}, \citenamefont {Foo}, \citenamefont {Fütterer}, \citenamefont {Gangadharan}, \citenamefont
  {Goodall}, \citenamefont {Gorelli}, \citenamefont {Goudeau}, \citenamefont {Guidoni}, \citenamefont {Guimiot}, \citenamefont {Haggerty}, \citenamefont {Hansen}, \citenamefont {Haque}, \citenamefont {Hillairet}, \citenamefont {Hoang}, \citenamefont {How}, \citenamefont {Huang}, \citenamefont {Humphrey}, \citenamefont {Isupova}, \citenamefont {Jeandet}, \citenamefont {Johnson}, \citenamefont {Jones}, \citenamefont {Kastek}, \citenamefont {Kent}, \citenamefont {Klima}, \citenamefont {Kulshrestha}, \citenamefont {Kumar}, \citenamefont {Kuszaj}, \citenamefont {Köhn-Seemann}, \citenamefont {Langendorf}, \citenamefont {Lanteri}, \citenamefont {Lee}, \citenamefont {Leonard}, \citenamefont {Lequette}, \citenamefont {Lim}, \citenamefont {Magarde}, \citenamefont {Malhotra}, \citenamefont {Martinelli}, \citenamefont {Masood}, \citenamefont {McHardy}, \citenamefont {Modi}, \citenamefont {Montes}, \citenamefont {Mumford}, \citenamefont {Munn}, \citenamefont {Murphy}, \citenamefont {Nie}, \citenamefont {Pannala},
  \citenamefont {Parashar}, \citenamefont {Patel}, \citenamefont {Pavon}, \citenamefont {Polak}, \citenamefont {Pérez}, \citenamefont {Qudsi}, \citenamefont {Raj}, \citenamefont {Rajashekar}, \citenamefont {Rao}, \citenamefont {Reep}, \citenamefont {Richardson}, \citenamefont {Roberts}, \citenamefont {Rojas~Zelaya}, \citenamefont {Salcido}, \citenamefont {Savcheva}, \citenamefont {Shen}, \citenamefont {Sheng}, \citenamefont {Sherpa}, \citenamefont {Schneck}, \citenamefont {Silvestri}, \citenamefont {Simon}, \citenamefont {Singh}, \citenamefont {Singh}, \citenamefont {Sipőcz}, \citenamefont {Skinner}, \citenamefont {Skrzypczak}, \citenamefont {Smirnov}, \citenamefont {Sobeske}, \citenamefont {Spedicato}, \citenamefont {Stansby}, \citenamefont {Stinson}, \citenamefont {Švancarová}, \citenamefont {Tavant}, \citenamefont {Tranquilino}, \citenamefont {Ulrich}, \citenamefont {Varnish}, \citenamefont {Vincena}, \citenamefont {Vo}, \citenamefont {Xu}, \citenamefont {Wu}, \citenamefont {Yip},\ and\ \citenamefont
  {Zhang}}]{plasmapy_community_2024_12788848}%
  \BibitemOpen
  \bibfield  {author} {\bibinfo {author} {\bibfnamefont {PlasmaPy}~\bibnamefont {Community}}, \bibinfo {author} {\bibfnamefont {N.~A.}\ \bibnamefont {Murphy}}, \bibinfo {author} {\bibfnamefont {E.}~\bibnamefont {Everson}}, \bibinfo {author} {\bibfnamefont {D.}~\bibnamefont {Stańczak-Marikin}}, \bibinfo {author} {\bibfnamefont {P.}~\bibnamefont {Heuer}}, \bibinfo {author} {\bibfnamefont {P.}~\bibnamefont {Kozlowski}}, \bibinfo {author} {\bibfnamefont {J.}~\bibnamefont {Addison}}, \bibinfo {author} {\bibfnamefont {A.~F.}\ \bibnamefont {Ahamed}}, \bibinfo {author} {\bibfnamefont {C.}~\bibnamefont {Arran}}, \bibinfo {author} {\bibfnamefont {H.}~\bibnamefont {Bagherianlemraski}}, \bibinfo {author} {\bibfnamefont {J.}~\bibnamefont {Beckers}}, \bibinfo {author} {\bibfnamefont {M.}~\bibnamefont {Bedmutha}}, \bibinfo {author} {\bibfnamefont {J.}~\bibnamefont {Bergeron}}, \bibinfo {author} {\bibfnamefont {L.}~\bibnamefont {Bessi}}, \bibinfo {author} {\bibfnamefont {R.}~\bibnamefont {Britten}}, \bibinfo {author}
  {\bibfnamefont {S.}~\bibnamefont {Brown}}, \bibinfo {author} {\bibfnamefont {K.}~\bibnamefont {Bryant}}, \bibinfo {author} {\bibfnamefont {S.}~\bibnamefont {Carroll}}, \bibinfo {author} {\bibfnamefont {C.}~\bibnamefont {Cartagena-Sanchez}}, \bibinfo {author} {\bibfnamefont {S.}~\bibnamefont {Chambers}}, \bibinfo {author} {\bibfnamefont {A.}~\bibnamefont {Chattopadhyay}}, \bibinfo {author} {\bibfnamefont {A.}~\bibnamefont {Choubey}}, \bibinfo {author} {\bibfnamefont {S.}~\bibnamefont {Choudhary}}, \bibinfo {author} {\bibfnamefont {C.}~\bibnamefont {Clauss}}, \bibinfo {author} {\bibfnamefont {J.}~\bibnamefont {Deal}}, \bibinfo {author} {\bibfnamefont {G.}~\bibnamefont {Decristoforo}}, \bibinfo {author} {\bibfnamefont {D.~A.}\ \bibnamefont {Diaz~Riega}}, \bibinfo {author} {\bibfnamefont {F.~M.}\ \bibnamefont {Dover}}, \bibinfo {author} {\bibfnamefont {D.}~\bibnamefont {Drozdov}}, \bibinfo {author} {\bibfnamefont {T.}~\bibnamefont {Du}}, \bibinfo {author} {\bibfnamefont {L.}~\bibnamefont {Einhorn}}, \bibinfo
  {author} {\bibfnamefont {T.}~\bibnamefont {Fan}}, \bibinfo {author} {\bibfnamefont {S.~I.}\ \bibnamefont {Farid}}, \bibinfo {author} {\bibfnamefont {M.}~\bibnamefont {Fischer}}, \bibinfo {author} {\bibfnamefont {B.}~\bibnamefont {Foo}}, \bibinfo {author} {\bibfnamefont {H.-A.}\ \bibnamefont {Fütterer}}, \bibinfo {author} {\bibfnamefont {R.}~\bibnamefont {Gangadharan}}, \bibinfo {author} {\bibfnamefont {B.}~\bibnamefont {Goodall}}, \bibinfo {author} {\bibfnamefont {M.}~\bibnamefont {Gorelli}}, \bibinfo {author} {\bibfnamefont {G.}~\bibnamefont {Goudeau}}, \bibinfo {author} {\bibfnamefont {S.}~\bibnamefont {Guidoni}}, \bibinfo {author} {\bibfnamefont {J.}~\bibnamefont {Guimiot}}, \bibinfo {author} {\bibfnamefont {C.}~\bibnamefont {Haggerty}}, \bibinfo {author} {\bibfnamefont {R.~S.}\ \bibnamefont {Hansen}}, \bibinfo {author} {\bibfnamefont {M.}~\bibnamefont {Haque}}, \bibinfo {author} {\bibfnamefont {J.}~\bibnamefont {Hillairet}}, \bibinfo {author} {\bibfnamefont {C.}~\bibnamefont {Hoang}}, \bibinfo {author}
  {\bibfnamefont {P.~Z.}\ \bibnamefont {How}}, \bibinfo {author} {\bibfnamefont {Y.-M.}\ \bibnamefont {Huang}}, \bibinfo {author} {\bibfnamefont {N.}~\bibnamefont {Humphrey}}, \bibinfo {author} {\bibfnamefont {M.}~\bibnamefont {Isupova}}, \bibinfo {author} {\bibfnamefont {A.}~\bibnamefont {Jeandet}}, \bibinfo {author} {\bibfnamefont {E.}~\bibnamefont {Johnson}}, \bibinfo {author} {\bibfnamefont {E.}~\bibnamefont {Jones}}, \bibinfo {author} {\bibfnamefont {M.}~\bibnamefont {Kastek}}, \bibinfo {author} {\bibfnamefont {J.}~\bibnamefont {Kent}}, \bibinfo {author} {\bibfnamefont {D.}~\bibnamefont {Klima}}, \bibinfo {author} {\bibfnamefont {S.}~\bibnamefont {Kulshrestha}}, \bibinfo {author} {\bibfnamefont {S.}~\bibnamefont {Kumar}}, \bibinfo {author} {\bibfnamefont {P.}~\bibnamefont {Kuszaj}}, \bibinfo {author} {\bibfnamefont {A.}~\bibnamefont {Köhn-Seemann}}, \bibinfo {author} {\bibfnamefont {S.}~\bibnamefont {Langendorf}}, \bibinfo {author} {\bibfnamefont {A.}~\bibnamefont {Lanteri}}, \bibinfo {author}
  {\bibfnamefont {T.~T.}\ \bibnamefont {Lee}}, \bibinfo {author} {\bibfnamefont {D.}~\bibnamefont {Leonard}}, \bibinfo {author} {\bibfnamefont {N.}~\bibnamefont {Lequette}}, \bibinfo {author} {\bibfnamefont {P.~L.}\ \bibnamefont {Lim}}, \bibinfo {author} {\bibfnamefont {A.}~\bibnamefont {Magarde}}, \bibinfo {author} {\bibfnamefont {R.}~\bibnamefont {Malhotra}}, \bibinfo {author} {\bibfnamefont {J.~V.}\ \bibnamefont {Martinelli}}, \bibinfo {author} {\bibfnamefont {M.}~\bibnamefont {Masood}}, \bibinfo {author} {\bibfnamefont {I.}~\bibnamefont {McHardy}}, \bibinfo {author} {\bibfnamefont {D.}~\bibnamefont {Modi}}, \bibinfo {author} {\bibfnamefont {K.}~\bibnamefont {Montes}}, \bibinfo {author} {\bibfnamefont {S.}~\bibnamefont {Mumford}}, \bibinfo {author} {\bibfnamefont {J.}~\bibnamefont {Munn}}, \bibinfo {author} {\bibfnamefont {L.}~\bibnamefont {Murphy}}, \bibinfo {author} {\bibfnamefont {S.}~\bibnamefont {Nie}}, \bibinfo {author} {\bibfnamefont {M.}~\bibnamefont {Pannala}}, \bibinfo {author} {\bibfnamefont
  {T.}~\bibnamefont {Parashar}}, \bibinfo {author} {\bibfnamefont {N.}~\bibnamefont {Patel}}, \bibinfo {author} {\bibfnamefont {F.~S.}\ \bibnamefont {Pavon}}, \bibinfo {author} {\bibfnamefont {J.}~\bibnamefont {Polak}}, \bibinfo {author} {\bibfnamefont {R.~D.}\ \bibnamefont {Pérez}}, \bibinfo {author} {\bibfnamefont {R.}~\bibnamefont {Qudsi}}, \bibinfo {author} {\bibfnamefont {R.}~\bibnamefont {Raj}}, \bibinfo {author} {\bibfnamefont {V.}~\bibnamefont {Rajashekar}}, \bibinfo {author} {\bibfnamefont {A.}~\bibnamefont {Rao}}, \bibinfo {author} {\bibfnamefont {J.}~\bibnamefont {Reep}}, \bibinfo {author} {\bibfnamefont {S.}~\bibnamefont {Richardson}}, \bibinfo {author} {\bibfnamefont {J.}~\bibnamefont {Roberts}}, \bibinfo {author} {\bibfnamefont {R.}~\bibnamefont {Rojas~Zelaya}}, \bibinfo {author} {\bibfnamefont {A.}~\bibnamefont {Salcido}}, \bibinfo {author} {\bibfnamefont {A.}~\bibnamefont {Savcheva}}, \bibinfo {author} {\bibfnamefont {C.}~\bibnamefont {Shen}}, \bibinfo {author} {\bibfnamefont
  {A.}~\bibnamefont {Sheng}}, \bibinfo {author} {\bibfnamefont {D.~N.}\ \bibnamefont {Sherpa}}, \bibinfo {author} {\bibfnamefont {C.}~\bibnamefont {Schneck}}, \bibinfo {author} {\bibfnamefont {L.}~\bibnamefont {Silvestri}}, \bibinfo {author} {\bibfnamefont {T.}~\bibnamefont {Simon}}, \bibinfo {author} {\bibfnamefont {A.}~\bibnamefont {Singh}}, \bibinfo {author} {\bibfnamefont {A.}~\bibnamefont {Singh}}, \bibinfo {author} {\bibfnamefont {B.}~\bibnamefont {Sipőcz}}, \bibinfo {author} {\bibfnamefont {C.}~\bibnamefont {Skinner}}, \bibinfo {author} {\bibfnamefont {T.~A.}\ \bibnamefont {Skrzypczak}}, \bibinfo {author} {\bibfnamefont {N.}~\bibnamefont {Smirnov}}, \bibinfo {author} {\bibfnamefont {S.}~\bibnamefont {Sobeske}}, \bibinfo {author} {\bibfnamefont {M.}~\bibnamefont {Spedicato}}, \bibinfo {author} {\bibfnamefont {D.}~\bibnamefont {Stansby}}, \bibinfo {author} {\bibfnamefont {T.}~\bibnamefont {Stinson}}, \bibinfo {author} {\bibfnamefont {M.}~\bibnamefont {Švancarová}}, \bibinfo {author} {\bibfnamefont
  {A.}~\bibnamefont {Tavant}}, \bibinfo {author} {\bibfnamefont {V.}~\bibnamefont {Tranquilino}}, \bibinfo {author} {\bibfnamefont {T.}~\bibnamefont {Ulrich}}, \bibinfo {author} {\bibfnamefont {T.}~\bibnamefont {Varnish}}, \bibinfo {author} {\bibfnamefont {S.}~\bibnamefont {Vincena}}, \bibinfo {author} {\bibfnamefont {T.}~\bibnamefont {Vo}}, \bibinfo {author} {\bibfnamefont {S.}~\bibnamefont {Xu}}, \bibinfo {author} {\bibfnamefont {T.}~\bibnamefont {Wu}}, \bibinfo {author} {\bibfnamefont {C.~H.}\ \bibnamefont {Yip}}, \ and\ \bibinfo {author} {\bibfnamefont {C.}~\bibnamefont {Zhang}},\ }\href {\doibase 10.5281/zenodo.12788848} {\enquote {\bibinfo {title} {Plasmapy},}\ } (\bibinfo {year} {2024})\BibitemShut {NoStop}%
\bibitem [{\citenamefont {Zhang}\ \emph {et~al.}(2025)\citenamefont {Zhang}, \citenamefont {Gao}, \citenamefont {Ji}, \citenamefont {Russell}, \citenamefont {Pomraning}, \citenamefont {Griff-McMahon}, \citenamefont {Klein}, \citenamefont {Kuranz},\ and\ \citenamefont {Wei}}]{zhang2025}%
  \BibitemOpen
  \bibfield  {author} {\bibinfo {author} {\bibfnamefont {Y.}~\bibnamefont {Zhang}}, \bibinfo {author} {\bibfnamefont {L.}~\bibnamefont {Gao}}, \bibinfo {author} {\bibfnamefont {H.}~\bibnamefont {Ji}}, \bibinfo {author} {\bibfnamefont {B.~K.}\ \bibnamefont {Russell}}, \bibinfo {author} {\bibfnamefont {G.}~\bibnamefont {Pomraning}}, \bibinfo {author} {\bibfnamefont {J.}~\bibnamefont {Griff-McMahon}}, \bibinfo {author} {\bibfnamefont {S.}~\bibnamefont {Klein}}, \bibinfo {author} {\bibfnamefont {C.}~\bibnamefont {Kuranz}}, \ and\ \bibinfo {author} {\bibfnamefont {M.}~\bibnamefont {Wei}},\ }\href {https://arxiv.org/abs/2505.02326} {\enquote {\bibinfo {title} {Diagnosing electric and magnetic fields in laser-driven coil targets},}\ } (\bibinfo {year} {2025}),\ \Eprint {http://arxiv.org/abs/2505.02326} {arXiv:2505.02326 [physics.plasm-ph]} \BibitemShut {NoStop}%
\bibitem [{\citenamefont {Yoon}\ \emph {et~al.}(2021)\citenamefont {Yoon}, \citenamefont {Kim}, \citenamefont {Choi}, \citenamefont {Sung}, \citenamefont {Lee}, \citenamefont {Lee},\ and\ \citenamefont {Nam}}]{Yoon_2021}%
  \BibitemOpen
  \bibfield  {author} {\bibinfo {author} {\bibfnamefont {J.~W.}\ \bibnamefont {Yoon}}, \bibinfo {author} {\bibfnamefont {Y.~G.}\ \bibnamefont {Kim}}, \bibinfo {author} {\bibfnamefont {I.~W.}\ \bibnamefont {Choi}}, \bibinfo {author} {\bibfnamefont {J.~H.}\ \bibnamefont {Sung}}, \bibinfo {author} {\bibfnamefont {H.~W.}\ \bibnamefont {Lee}}, \bibinfo {author} {\bibfnamefont {S.~K.}\ \bibnamefont {Lee}}, \ and\ \bibinfo {author} {\bibfnamefont {C.~H.}\ \bibnamefont {Nam}},\ }\href {\doibase 10.1364/OPTICA.420520} {\bibfield  {journal} {\bibinfo  {journal} {Optica}\ }\textbf {\bibinfo {volume} {8}},\ \bibinfo {pages} {630} (\bibinfo {year} {2021})}\BibitemShut {NoStop}%
\bibitem [{\citenamefont {Piazza}, \citenamefont {Willingale},\ and\ \citenamefont {Zuegel}(2022)}]{MP3_workshop}%
  \BibitemOpen
  \bibfield  {author} {\bibinfo {author} {\bibfnamefont {A.~D.}\ \bibnamefont {Piazza}}, \bibinfo {author} {\bibfnamefont {L.}~\bibnamefont {Willingale}}, \ and\ \bibinfo {author} {\bibfnamefont {J.~D.}\ \bibnamefont {Zuegel}},\ }\href {https://arxiv.org/abs/2211.13187} {\enquote {\bibinfo {title} {Multi-petawatt physics prioritization (mp3) workshop report},}\ } (\bibinfo {year} {2022}),\ \Eprint {http://arxiv.org/abs/2211.13187} {arXiv:2211.13187 [hep-ph]} \BibitemShut {NoStop}%
\end{thebibliography}
%

\end{document}